\documentclass[preprint,aps,prd,showpacs,superscriptaddress,nofootinbib,tightenlines]{revtex4}
\usepackage{amsfonts}
\usepackage{multirow}
\usepackage{mathrsfs}
\usepackage{graphicx}
\usepackage{amsmath}
\usepackage{amssymb}
\usepackage{bm}
\usepackage{bbm}
\usepackage{color}
\usepackage{array}
\usepackage{diagbox}
\usepackage{float}
\usepackage{upgreek}
\usepackage{subcaption}
\usepackage{diagbox}

\def\OMIT#1{}

\def\hlinew#1{%
  \noalign{\ifnum0=`}\fi\hrule \@height #1 \futurelet
   \reserved@a\@xhline}
\usepackage{array}
\newcommand{\PreserveBackslash}[1]{\let\temp=\\#1\let\\=\temp}
\newcolumntype{C}[1]{>{\PreserveBackslash\centering}p{#1}}
\newcolumntype{R}[1]{>{\PreserveBackslash\raggedleft}p{#1}}
\newcolumntype{L}[1]{>{\PreserveBackslash\raggedright}p{#1}}

\newcommand{\nn}{\nonumber}

\newcommand{\beq}{\begin{equation}}
\newcommand{\eeq}{\end{equation}}
\newcommand{\bqa}{\begin{eqnarray}}
\newcommand{\eqa}{\end{eqnarray}}

\newcommand{\jpsi}{J/\psi}

\newcommand\fverb{\setbox\fverbbox=\hbox\bgroup\verb}
\newcommand\fverbdo{\egroup\medskip\noindent%
			\fbox{\unhbox\fverbbox}\ }
\newcommand\fverbit{\egroup\item[\fbox{\unhbox\fverbbox}]}
\newbox\fverbbox

\newcommand{\Lmu}{\ln\dfrac{\mu_\Lambda^2}{m_Q^2}}
\newcommand{\tabincell}[2]{\begin{tabular}{@{}#1@{}}#2\end{tabular}}

\makeatletter

\newcommand{\Rmnum}[1]{\expandafter\@slowromancap\romannumeral #1@}

\usepackage{tikz}
\usetikzlibrary {shapes.misc}
\usetikzlibrary{arrows.meta}
\usetikzlibrary{calc}
\usetikzlibrary{
  decorations.pathmorphing,
  decorations.pathreplacing,
  decorations.markings
}
\usetikzlibrary{patterns}

\tikzset{
  every picture/.style={semithick, line cap=round},
  scalar/.style={dashed},
  fermion/.default=0.5,
  fermion/.style={postaction={decorate, decoration={
    markings,
    mark=at position #1 with {\arrow{Stealth[angle=30:7pt,inset=1.5pt]}},
    transform={xshift={3.5pt*cos(15)}}
  }}},
  antifermion/.default=0.5,
  antifermion/.style={postaction={decorate, decoration={
    markings,
    mark=at position #1 with {\arrowreversed{Stealth[angle=30:7pt,inset=1.5pt]}},
    transform={xshift={-3.5pt*cos(15)}}
  }}},
  gluon/.default=3pt,
  gluon/.style={decorate, decoration={
    coil,
    amplitude=0.5*#1,
    aspect=1,
    segment length=#1
  }},
  rgluon/.default=3pt,
  rgluon/.style={decorate, decoration={
    coil,
    amplitude=-0.5*#1,
    aspect=-1,
    segment length=#1
  }},
  gluonpre/.default=0pt,
  gluonpre/.style={decorate, decoration={
    coil,
    amplitude=1.5pt,
    aspect=1,
    segment length=3pt,
    pre length=#1
  }},
  rgluonpre/.default=0pt,
  rgluonpre/.style={decorate, decoration={
    coil,
    amplitude=-1.5pt,
    aspect=-1,
    segment length=3pt,
    pre length=#1
  }},
  crossmark/.style={cross out, draw=red, inner sep=2pt},
  counter/.style={path picture={
    \draw (path picture bounding box.south east) --
      (path picture bounding box.north west)
      (path picture bounding box.south west) --
      (path picture bounding box.north east);
  }},
  cnode/.default=8pt,
  cnode/.style={inner sep=0pt, minimum size=#1, circle}
}

\makeatother

\usepackage{babel}
\begin{document}
%%%%%%%%%%%%%%%%%%%%%%%%%%%%%%%%%%%%%%%%%%%%%%%%%%%%%%%%%%%%%%%%%%%%%%%%%%%%%%
\title{\mbox{}\\[10pt]
Complete three-loop QCD corrections to leptonic width of vector quarkonium}
%%%%%%%%%%%%%%%%%%%%%%%%%%%%%%%%%%%%%%%%%%%%%%%%%%%%%%%%%%%%%%%%%%%%%%%%%%%%%%

\author{Feng Feng\footnote{f.feng@outlook.com}}
\affiliation{China University of Mining and Technology, Beijing 100083, China\vspace{0.2cm}}
\affiliation{Institute of High Energy Physics, Chinese Academy of
Sciences, Beijing 100049, China\vspace{0.2cm}}

\author{Yu Jia\footnote{jiay@ihep.ac.cn}}
\affiliation{Institute of High Energy Physics, Chinese Academy of
Sciences, Beijing 100049, China\vspace{0.2cm}}
\affiliation{School of Physics, University of Chinese Academy of Sciences, Beijing 100049, China\vspace{0.2cm}}

\author{Zhewen Mo\footnote{mozw@ihep.ac.cn }}
\affiliation{Institute of High Energy Physics, Chinese Academy of
Sciences, Beijing 100049, China\vspace{0.2cm}}
\affiliation{School of Physics, University of Chinese Academy of Sciences, Beijing 100049, China\vspace{0.2cm}}

\author{Jichen Pan\footnote{panjichen@ihep.ac.cn}}
\affiliation{Institute of High Energy Physics, Chinese Academy of
Sciences, Beijing 100049, China\vspace{0.2cm}}
\affiliation{School of Physics, University of Chinese Academy of Sciences, Beijing 100049, China\vspace{0.2cm}}

\author{Wen-Long Sang~\footnote{wlsang@swu.edu.cn}}
 \affiliation{School of Physical Science and Technology, Southwest University, Chongqing 400700, China\vspace{0.2cm}}

\author{Jia-Yue Zhang\footnote{zhangjiayue@ihep.ac.cn}}
\affiliation{Institute of High Energy Physics, Chinese Academy of
Sciences, Beijing 100049, China\vspace{0.2cm}}
\affiliation{School of Physics, University of Chinese Academy of Sciences, Beijing 100049, China\vspace{0.2cm}}

\date{\today}
%%%%%%%%%%%%%%%%%%%%%%%%%%%%%%%%%%%%%%%%%%%%%%%%%%%%%%%%%%%%%%%%%%%%%%%%%%%%%%
\begin{abstract}
Within the nonrelativistic QCD (NRQCD) factorization framework, we compute the order-$\alpha_s^3$ perturbative corrections to the leptonic decay of the $\Upsilon$ and
$\jpsi$ with high numerical accuracy, at the lowest order in velocity expansion. We confirm the existing three-loop results in literature.
Furthermore, we explicitly consider the complex-valued light-by-light (singlet) contributions. We for the first time also
consider the finite charm quark mass effect for $\Upsilon$ leptonic decay, and obtain a new piece of 3-loop contribution to the anomalous dimension related to
the composite NRQCD bilinear of vector current that arises from the nonzero charm quark mass.
Based on the complete three-loop NRQCD short-distance coefficients, we also present a comprehensive phenomenological analysis for $\Upsilon(J/\psi)$ leptonic width.
\end{abstract}

\maketitle

\section{Introduction}

The leptonic decay of vector quarkonium (exemplified by $J/\psi,\Upsilon\to l^+l^-$), as the simplest quarkonium decay process,
has already been extensively studied both experimentally and theoretically. On the experimental side, these very clean
decay channels have already been measured with very high precision. These decay channels are also very useful to reconstruct the vector quarkonium in various
collision experiments. On the theoretically ground, these types of quarkonium electromagnetic decay processes directly probe the vector quarkonium decay constant,
a basic nonperturbative parameter characterizing the quarkonium dynamics. Numerous theoretical efforts have been devoted to predicting the quakonium decay constant,
including all sorts of phenomenological model predictions together with the first-principle approach, the lattice QCD simulation~\cite{Hatton:2020qhk,Hatton:2021dvg}.
Reassuringly, the good agreement is reached between the experimental measurement and lattice result.

From the theoretical angle, nonrelativistic QCD (NRQCD)~\cite{Caswell:1985ui,Bodwin:1994jh} also provides a model-indepent framework to investigate the vector quarkonium leptonic decay. Since the leading-order (LO) prediction has been known since $1967$~\cite{VanRoyen:1967nq}, a tremendous progress including higher-order corrections within the NRQCD framework has been constantly made during the past half century. For example, the $\mathcal{O}(\alpha_s v^0)$
correction was computed in late 70s~\cite{Barbieri:1975ki,Celmaster:1978yz} ($v$ denotes the typical heavy quark velocity inside quarkonium), while
the leading relativistic correction of $\mathcal{O}(\alpha_s^0v^2)$ and $\mathcal{O}(\alpha_s^0v^4)$ were investigated in \cite{Bodwin:1994jh,Bodwin:2002cfe}.
The $\mathcal{O}(\alpha_s v^2)$ correction is calculated in \cite{Luke:1997ys}.
The two-loop QCD corrections yet at lowest order in $v$ were calculated in \cite{Czarnecki:1997vz,Beneke:1997jm,Kniehl:2006qw,Egner:2021lxd}.
Finally, the $\mathcal{O}(\alpha_s^3)$ fermionic~\cite{Marquard:2006qi,Marquard:2009bj} and
and pure gluonic corrections~\cite{Marquard:2014pea,Beneke:2014qea} to $\Upsilon\to l^+l^-$ have also been known in the past decade.

 In this paper, we make some further progress for the three-loop matching of the vector current into QCD with the corresponding operator in NRQCD. Explicitly speaking,
 we refine the existing three-loop results by including the light-by-light (singlet) contributions, as well as including the finite charm mass effect in $\Upsilon$ decay.
 There are some nontrivial technical challenges we have to overcome, because we have to deal with complex-valued loop integrals for the former, and deal with two mass scales
 in loop integration for the latter. Our calculation is made possible with the aid of the newly developed auxiliary mass flow (AMF) method to compute the master integrals with exceptional numerical accuracy~\cite{Liu:2017jxz,Liu:2020kpc,Liu:2022chg}. An interesting theoretical progress is that we find that the finite charm mass effect leads to
 a new piece of contribution to the three-loop anomalous dimension for the NRQCD vector current.
 We also present a comprehensive analysis of the $\jpsi$ and $\Upsilon$ leptonic decay width at the three-loop level, carefully assessing
 various sources of theoretical uncertainty.

The rest of the paper is structured as follows.  In Section~\ref{sec:formalism}, we recapitulate the general formalism for NRQCD factorization and vector current matching. In Section~\ref{sec:calculation} we sketch the strategy of our three-loop calculation and present the numerical SDCs as well as the analytical
expressions of the renormalization constant and the anomalous dimension of the NRQCD vector current. Section~\ref{sec:phenomenology} is devoted to
phenomenological analysis of the confrontation of the finest NRQCD predictions to the measured $\jpsi$ and $\Upsilon$ leptonic decay width.
Finally, we summarize Section~\ref{sec:summary}.
%-------------------------------
%--------------------------------

\section{Matching of the electromagnetic current \label{sec:formalism}}

We start by define the leptonic decay constant $f_V$ for a given vector quarkonium $V$, through the vacuum-to-quarkonium matrix element of the electromagnetic current $\mathcal{J}_\text{EM}^\mu$:
%--------------------------------
\begin{align}
%--------------------------------
&\left\langle 0 \vert \mathcal{J}_\text{EM}^\mu \vert V(\boldsymbol\upepsilon)\right\rangle={M_V f_V\varepsilon_V^\mu},\nn\\
&\mathcal{J}_\text{EM}^\mu=\sum_fe_f\bar{\Psi}_f\gamma^\mu\Psi_f,
\label{em:current:decay:constant}
%--------------------------------
\end{align}
%--------------------------------
where $M_V$ and $\varepsilon_V^\mu$ denote the mass and polarization vector of a vector quarkonium. The summation in definition of the EM current
includes all the flavors of quarks, with $e_u=2/3$ for the up-type quarks and $e_d=-1/3$ for down-type quarks.

The leptonic decay width of the vector quarkonium $V$ then becomes
\begin{align}
  \Gamma(V\rightarrow l^+l^-)={\dfrac{4\uppi\alpha^2 }{3M_V}|f_V|^2},
\end{align}
where $\alpha$ is the QED fine structure constant.

According to NRQCD factorization formula, the decay constant $f_V$ is not a completely nonperturvative object.
At the lowest order in velocity expansion, it can be further factorized in the following form:
%--------------------------------
\begin{align}
  f_V\epsilon_V^i=&
  {\sqrt{2M_V}e_Q\mathcal{C}\dfrac{\left\langle 0 \vert \tilde{j}^i\vert V(\boldsymbol\upepsilon)\right\rangle}{M_V}}+\mathcal{O}\left(v^2\right)\nn\\
  =&{\sqrt{2M_V}e_Q\left(\mathcal{C}_\text{dir}+\sum_{f\neq Q}\mathcal{C}_\text{ind,f}\dfrac{e_f}{e_Q}\right)\dfrac{\left\langle 0 \vert \tilde{j}^i\vert V(\boldsymbol\upepsilon)\right\rangle}{M_V}}+\mathcal{O}\left(v^2\right),
%--------------------------------
\label{eq:NRQCD factorization}
%--------------------------------
\end{align}
%--------------------------------
where $\tilde{j}=\chi^\dagger\boldsymbol{\upsigma}\psi$ is the corresponding NRQCD vector currentoperator,
with $\sigma^i$ ($i=1,2,3$) denoting the Pauli matrices, and $\psi$ and $\chi^\dagger$ representing two-component spinor fields
that annihilate a heavy quark and a heavy anti-quark, respectively.
The factor $\sqrt{2M_V}$ has been explicitly inserted in the right-hand side of \eqref{eq:NRQCD factorization}, in order to
compensate the fact that the quarkonium state in the QCD side is relativistically normalized, where the quarkonium state in
the NRQCD matrix element is conventionally nonrelativistically normalized.
The dimensionless coefficient $\mathcal{C}$ is the short-distance coefficient (SDC), which encodes the effect from the relativistic quantum fluctuation,
which can be reliably computed in perturbation theory owing to asymptotic freedom of QCD.
For the sake of clarity, we devide the SDCs into two categories, the {\it direct} one and the {\it indirect} one. The former corresponds to the matching of heavy-quark vector current $j^\mu=\bar{\Psi}_Q\gamma^\mu\Psi_Q$ (the quark flavor in the current is the same as with the leading Fock state of a vector quarkonium). The latter arises from the contribution from the the quark flavors in the EM current in (\ref{em:current:decay:constant}) which differ from the dominant quark flavor comprising the vector quarkonium. This can be viewed as the the manifestation of high-order Fock state components of a vector quarkonium with a short-distance origin.
For example, the $\Upsilon$ may contain a tiny yet nonzero content of $u\bar{u}({}^3S_1)$ or $c\bar{c}({}^3S_1)$ components, since $b\bar{b}({}^3S_1)$ might mix with these states through three-gluon annihilation.

The SDC can be systematically computed in perturbation theory since $m_Q\gg\Lambda_\text{QCD}$.
The main purpose of this paper is to complete the evaluation of SDCs through ${\cal O}(\alpha_s^3)$, {\it}, include the {\it indirect} contributions (often referred to as
singlet or light-by-light contribution) and include the diagrams where the quarks in the closed loop can carry nonvanishing mass $m_M\neq m_Q$.

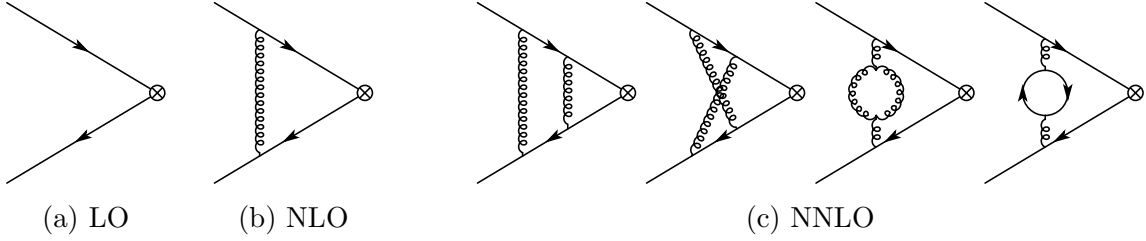
\begin{figure}[t]
  \begin{subfigure}{.16\textwidth}
     \begin{tikzpicture}[baseline]
    \coordinate (I1) at (-2, 1.2);
    \coordinate (I2) at (-2, -1.2);
    \coordinate (O) at (0,0);

    \draw[fermion] (I1) -- (O);
    \draw[antifermion] (I2) --  (O);
    \fill[counter, fill=white, draw] (O) circle (0.1);
  \end{tikzpicture}
  \caption{LO}
  \end{subfigure}
  \begin{subfigure}{.16\textwidth}
    \begin{tikzpicture}[baseline]
      \coordinate (I1) at (-2, 1.2);
      \coordinate (I2) at (-2, -1.2);
      \coordinate (O) at (0,0);

      \draw[fermion] (I1) -- node[pos=0.3](V1){}(O);
      \draw[antifermion] (I2) -- node[pos=0.3](V2){} (O);
      \fill[counter, fill=white, draw] (O) circle (0.1);
      \draw[rgluon] (V1.center) -- (V2.center);
    \end{tikzpicture}
    \caption{NLO}
  \end{subfigure}
  \begin{subfigure}{.66\textwidth}
    \begin{tikzpicture}[baseline]
      \coordinate (I1) at (-2, 1.2);
      \coordinate (I2) at (-2, -1.2);
      \coordinate (O) at (0,0);

      \draw[fermion] (I1) -- node[pos=0.3](V1){} node[pos=0.6](V2){} (O);
      \draw[antifermion] (I2) -- node[pos=0.3](V3){} node[pos=0.6](V4){} (O);
      \fill[counter, fill=white, draw] (O) circle (0.1);
      \draw[rgluon] (V1.center) -- (V3.center);
      \draw[rgluon] (V2.center) -- (V4.center);
    \end{tikzpicture}
    \begin{tikzpicture}[baseline]
      \coordinate (I1) at (-2, 1.2);
      \coordinate (I2) at (-2, -1.2);
      \coordinate (O) at (0,0);

      \draw[fermion] (I1) -- node[pos=0.3](V1){} node[pos=0.6](V2){} (O);
      \draw[antifermion] (I2) -- node[pos=0.3](V3){} node[pos=0.6](V4){} (O);
      \fill[counter, fill=white, draw] (O) circle (0.1);
      \draw[rgluon] (V1.center) -- (V4.center);
      \draw[rgluon] (V2.center) -- (V3.center);
    \end{tikzpicture}
    \begin{tikzpicture}[baseline]
      \coordinate (I1) at (-2,1.2);
      \coordinate (I2) at (-2,-1.2);
      \coordinate (O) at (0,0);

      \draw[fermion] (I1) -- node[pos=0.4](V1){} (O);
      \draw[antifermion] (I2) -- node[pos=0.4](V2){} (O);
      \fill[counter, fill=white, draw] (O) circle (0.1);

      \coordinate (V3) at (-1.2, 0.3);
      \coordinate (V4) at (-1.2, -0.3);

      \draw[rgluon] (V1.center) -- (V3);
      \draw[gluon] (V2.center) -- (V4);
      \draw[gluon=3.1pt] (V3) arc(90:-90:0.3);
      \draw[gluon=3.1pt] (V4) arc(-90:-270:0.3);
    \end{tikzpicture}
    \begin{tikzpicture}[baseline]
      \coordinate (I1) at (-2,1.2);
      \coordinate (I2) at (-2,-1.2);
      \coordinate (O) at (0,0);

      \draw[fermion] (I1) -- node[pos=0.4](V1){} (O);
      \draw[antifermion] (I2) -- node[pos=0.4](V2){} (O);
      \fill[counter, fill=white, draw] (O) circle (0.1);

      \coordinate (V3) at (-1.2,0.3);
      \coordinate (V4) at (-1.2,-0.3);

      \draw[rgluon] (V1.center) -- (V3);
      \draw[gluon] (V2.center) -- (V4);
      \draw[fermion] (V3) arc(90:-90:0.3); \draw[fermion] (V4) arc(-90:-270:0.3) 
      ;
    \end{tikzpicture}
    \caption{NNLO}
  \end{subfigure}
  \caption{Some representative Feynman diagrams up to two-loop order. The cross denotes the insertion of the external EM current. \label{fig:feynmandiagrams-NNLO}}
\end{figure}

The SDC can be determined through the standard perturbative matching procedure. One calculates the on-shell vertex functions in both perturbative QCD and perturbative
NRQCD sides, then solve the SDC through the following matching condition:
%--------------------------------
\begin{align}
%--------------------------------
  Z_2 \Gamma =&{\sqrt{2M_V}}\mathcal{C}\tilde{Z}_2 \tilde{Z}_v^{-1} \tilde{\Gamma}
  + \mathcal{O}\left(v^2\right),
\label{matching:equation}
%--------------------------------
\end{align}
%--------------------------------
where the quantities with a tilde are defined in NRQCD. $Z_2$ denotes the on-shell field strength renormalization
constant of the heavy quark. $\Gamma$ denotes the one-particle irreducible vertex diagrams
with on-shell heavy quarks. Some representative diagrams up to two-loop order are shown in FIG.~\ref{fig:feynmandiagrams-NNLO}.
${Z}_v=1$ since the vector current in QCD is a conserving current which does not require operator renormalization.
 $\tilde{Z}_v^{-1}$ is the renormalization constant affiliated with the NRQCD vector current
 $\tilde{\mathbf{j}}$.
 As for $\tilde{Z}_v$, the known analytical results only include the contributions of internal closed quark loops which are either massless or have the same mass with the
 external quark ($m_Q$)~\cite{Beneke:1997jm,Marquard:2006qi,Kniehl:2002yv,Beneke:2007pj}.
 If we extend to the case where the closed quark loop could have a different mass ($m_M$), it is possible that $\tilde{Z}_v$ might contain an extra unknown term denoted as $\gamma_0$:
%--------------------------------
\begin{align}
%--------------------------------
  \tilde{Z}_{v}= & 1+\left(\dfrac{\alpha_{s}(\mu_{\Lambda})}{\uppi}\right)^{2}\dfrac{C_{F}\uppi^{2}}{\epsilon}\left(\dfrac{1}{12}C_{F}+\dfrac{1}{8}C_{A}\right)+\left(\dfrac{\alpha_{s}(\mu_{\Lambda})}{\uppi}\right)^{3}C_{F}\uppi^{2}\nn \\
   & \times\Bigg\{ C_{F}^{2}\left[\dfrac{5}{144\epsilon^{2}}+\left(\dfrac{43}{144}-\dfrac{1}{2}\ln 2+\dfrac{5}{48}\Lmu\right)\dfrac{1}{\epsilon}\right]\nn \\
   & +C_{F}C_{A}\left[\dfrac{1}{864\epsilon^{2}}+\left(\dfrac{113}{324}+\dfrac{1}{4}\ln 2+\dfrac{5}{32}\Lmu\right)\dfrac{1}{\epsilon}\right]\nn \\
   & +C_{A}^{2}\left[-\dfrac{1}{16\epsilon^{2}}+\left(\dfrac{2}{27}+\dfrac{1}{4}\ln 2+\dfrac{1}{24}\Lmu\right)\dfrac{1}{\epsilon}\right]\nn \\
   & +T_{F}n_{L}\left[C_{F}\left(\dfrac{1}{54\epsilon^{2}}-\dfrac{25}{324\epsilon}\right)+C_{A}\left(\dfrac{1}{36\epsilon^{2}}-\dfrac{37}{432\epsilon}\right)\right]\nn \\
   & +T_{F}n_{H}\dfrac{C_{F}}{60\epsilon} +T_{F}n_{M}\dfrac{1}{\epsilon}\left[\gamma_{0}-\left(\dfrac{C_{F}}{18}+\dfrac{C_{A}}{12}\right){\ln\dfrac{\mu_{\Lambda}^{2}}{m_{Q}^{2}}}\right]\Bigg\}+{\cal O}\left(\alpha_{s}^{4}\right),
%--------------------------------
\end{align}
%--------------------------------
where $\mu_\Lambda$ is the NRQCD factorization scale, whose maximum value is around the heavy quark mass, the natural UV cutoff of NRQCD.
The group-theoretical factors are $T_F = 1/2, C_F=(N_c^2-1)/(2N_c)$, $C_A = N_c$ relevant for $SU(N_c)$.
$n_L=3$ is the number of light quark flavours, $n_H=1$ is the number of heavy quark. Here we introduce $n_M$ indicating that the number of
massive quark appearing the closed quark loop. For example, we choose $n_M=0$ in $J/\psi$ leptonic decay, but choose $n_M=1$ for $\Upsilon$
leptonic decay if we include the closed loop formed by the massive charm quark.
Note that the strong coupling constant in this paper is defined in the effective theory of QCD with $n_f=n_L+n_M$ active quark flavors unless otherwise specified.
$\tilde{Z}_v$ should return to the known single-mass-scale results with $n_H=2$ for $n_M=n_H$, \textit{i.e.}, $\gamma_0(m_M=m_Q)=C_F/60$~\cite{Marquard:2014pea}.

To expedite the matching procedure, we take the shortcut by neglecting the relative motion between
the external heavy quark and heavy antiquark when computing the vertex function $\Gamma$.
This amounts to applying the strategy of region~\cite{Beneke:1997zp} to directly extract the hard region contributions.
It saves lots of labors since there is no need to evaluate any loop diagrams in the NRQCD side.
As a consequence, one can simply set $\tilde{\Gamma}=1$ and $\tilde{Z}_2=1$ in (\ref{matching:equation}).

At the $\text{N}^3$LO in $\alpha_s$, the dimensionless SDCs are expected to bear the following structure:
\begin{align}
  \mathcal{C}\left(\dfrac{\mu_R}{m_Q},\dfrac{\mu_\Lambda}{m_Q},x\right)=&1+\dfrac{\alpha_{s}\left(\mu_{R}\right)}{\uppi}\mathcal{C}^{\left(1\right)}
  +\left[\dfrac{\alpha_{s}\left(\mu_{R}\right)}{\uppi}\right]^{2}\left[\mathcal{C}^{\left(1\right)}\dfrac{\beta_{0}}{4}\ln\dfrac{\mu_{R}^{2}}{m_Q^{2}}+\gamma_{v}^{\left(2\right)}\Lmu+\mathcal{C}^{\left(2\right)}\left(x\right)\right]\nn\\
  &+\left[\dfrac{\alpha_{s}\left(\mu_{R}\right)}{\uppi}\right]^{3}
  \Bigg\{
    \dfrac{\mathcal{C}^{\left(1\right)}}{16}\beta_{0}^{2}\ln^{2}\dfrac{\mu_{R}^{2}}{m_Q^{2}}
    +\left[\dfrac{\mathcal{C}^{\left(1\right)}}{16}\beta_{1} +\mathcal{C}^{\left(2\right)}\left(x\right)\dfrac{\beta_{0}}{2}\right]\ln\dfrac{\mu_{R}^{2}}{m_Q^{2}}\nn\\
    &+\gamma_{v}^{\left(2\right)}\dfrac{\beta_{0}}{2}\Lmu\ln\dfrac{\mu_{R}^{2}}{m_Q^{2}}
    +\dfrac{1}{4}\left[2\dfrac{\mathrm{d}\gamma_v^{(3)}\left(\mu_\Lambda\right)}{\mathrm{d}\ln\mu_\Lambda^2}-\beta_{0}\gamma_{v}^{\left(2\right)}\right]\ln^{2}\dfrac{\mu_{\Lambda}^{2}}{m_Q^{2}}\nn\\
    &+\left[\mathcal{C}^{\left(1\right)}\gamma_{v}^{\left(2\right)}+\gamma_{v}^{\left(3\right)}\left(m_Q\right)\right]\Lmu+\mathcal{C}^{\left(3\right)}\left(x\right)\Bigg\}+\mathcal{O}\left(\alpha_{s}^{4}\right),
  \end{align}
where $\mu_R$ is the renormalization scale. $\beta_0= 11C_A/3-4T_F n_f/3$ and $\beta_1=34C_A^2/3-20C_AT_F n_f/3-4C_FT_F n_f$ are the first two coefficients in the QCD $\beta$ function. Note $\mu_R$ has quite different physical origin from the NRQCD factorization scale $\mu_\Lambda$.
The mass ratio is defined by
\begin{align}
  x=\dfrac{m_M}{m_Q}.
\end{align}

$\gamma_{v}^{\left(2\right)}$ and $\gamma_{v}^{\left(3\right)}$ are the various coefficients of the anomalous dimension associated with the NRQCD vector current
$\tilde{\mathbf{j}}$, which can be deduced by taking the logarithmic derivative of the renormalization constant $\tilde{Z}_v$:
\begin{align}
  \gamma_v=&\dfrac{\mathrm{d}\ln\tilde{Z}_v}{\mathrm{d}\ln\mu_\Lambda^2}=\left(\dfrac{\alpha_s}{\uppi}\right)^2\gamma_v^{(2)}+\left(\dfrac{\alpha_s}{\uppi}\right)^3\gamma_v^{(3)}{\left(\mu_\Lambda\right)}+\mathcal{O}\left(\alpha_s^4\right).
\end{align}

The one- and two-loop corrections to the SDCs have been known long ago~\cite{,Kallen:1955fb,Czarnecki:1997vz,Beneke:1997jm,Kniehl:2006qw,Egner:2021lxd}:
\begin{align}
  \mathcal{C}^{(1)}=&-2C_F,\nn\\
  \mathcal{C}^{(2)}=&\left[-\dfrac{151}{72}
    +\dfrac{89\uppi^2}{144}
    -\dfrac{5\uppi^2}{6}\ln2-\dfrac{13}{4}\zeta(3)\right]C_AC_F
  +\left[\dfrac{23}{8}-\dfrac{79\uppi^2}{36}
    +\uppi^2\ln2-\dfrac{1}{2}\zeta(3)\right]C_F^2
  \nn\\
  &+\left(\dfrac{22}{9}-\dfrac{2\uppi^2}{9}\right)C_FT_Fn_H
  +\dfrac{11}{18}C_FT_Fn_L\nn\\
    &+n_MC_FT_F\bigg\{\dfrac{71}{72}+\dfrac{35x^2}{24}+\uppi^2\left(\dfrac{3}{32x}-\dfrac{11x}{48}-\dfrac{17x^3}{32}+\dfrac{2x^4}{9}\right)+\dfrac{1}{24}\left(23+19x^2\right)H_0(x)\nn\\
    &+\dfrac{4}{3}x^4H_0^2(x)+\left(\dfrac{3}{16x}-\dfrac{11x}{24}-\dfrac{17x^3}{16}+\dfrac{4x^4}{3}\right)\left[H_0(x)H_1(x)-H_{0,1}(x)\right]\nn\\
    &+\left(\dfrac{3}{16x}-\dfrac{11x}{24}-\dfrac{17x^3}{16}-\dfrac{4x^4}{3}\right)H_{-1,0}(x)+{\dfrac{2}{3}\ln\dfrac{m_Q^2}{m_M^2}}\bigg\},
\end{align}
where $H_*(x)$ are harmonic polylogarithms (HPLs).

\section{The SDCs in three loop\label{sec:calculation}}

\begin{figure}
  \begin{subfigure}{.22\textwidth}
  \centering
    \begin{tikzpicture}[baseline]
      \coordinate (I1) at (-2, 1.2);
      \coordinate (I2) at (-2, -1.2);
      \coordinate (O) at (0,0);
      \coordinate (V5) at (-1, 0);

      \draw[fermion] (I1) -- node[pos=0.2](V1){} node[pos=0.45](V2){} node[pos=0.7](V3){} (O);
      \draw[antifermion] (I2) -- node[pos=0.2](V4){} node[pos=0.45](V5){} node[pos=0.7](V6){} (O);
      \fill[counter, fill=white, draw] (O) circle (0.1);

      \draw[rgluon] (V1.center) -- (V4.center);
      \draw[rgluon] (V2.center) -- (V5.center);
      \draw[rgluon] (V3.center) -- (V6.center);
    \end{tikzpicture}
    \caption{$C_F^3$}
  \end{subfigure}
  \begin{subfigure}{.22\textwidth}
  \centering
    \begin{tikzpicture}[baseline]
      \coordinate (I1) at (-2, 1.2);
      \coordinate (I2) at (-2, -1.2);
      \coordinate (O) at (0,0);
      \coordinate (V5) at (-1, 0);

      \draw[fermion] (I1) -- node[pos=0.2](V1){} node[pos=1-0.8*2/3](V2){} node[pos=0.7](V3){} (O);
      \draw[antifermion] (I2) -- node[pos=0.15](V4){} node[pos=0.4](V5){} node[pos=0.6](V6){} (O);
      \fill[counter, fill=white, draw] (O) circle (0.1);

      \draw[rgluon] (V1.center) -- (V5.center);
      \draw[rgluon] (V2.center) -- (V6.center);
      \draw[rgluon] (V3.center) -- (V4.center);
    \end{tikzpicture}
    \caption{$(C_A-2C_F)^2C_F$}
  \end{subfigure}
  \begin{subfigure}{.22\textwidth}
  \centering
    \begin{tikzpicture}[baseline]
      \coordinate (I1) at (-2, 1.2);
      \coordinate (I2) at (-2, -1.2);
      \coordinate (O) at (0,0);
      \coordinate (V5) at (-1, 0);

      \draw[fermion] (I1) -- node[pos=0.2](V1){} node[pos=0.35](V2){} node[pos=0.7](V3){} (O);
      \draw[antifermion] (I2) -- node[pos=0.2](V4){} node[pos=0.35](V5){} node[pos=0.7](V6){} (O);
      \fill[counter, fill=white, draw] (O) circle (0.1);

      \draw[rgluon] (V1.center) -- (V4.center);
      \draw[rgluon] (V2.center) -- (V6.center);
      \draw[rgluon] (V3.center) -- (V5.center);
    \end{tikzpicture}
      \caption{$(C_A-2C_F)C_F^2$}
  \end{subfigure}
  \begin{subfigure}{.3\textwidth}
  \centering
    \begin{tikzpicture}[baseline]
      \coordinate (I1) at (-2, 1.2);
      \coordinate (I2) at (-2, -1.2);
      \coordinate (O) at (0,0);

      \draw[fermion] (I1) -- node[pos=0.2](V5){} node[pos=0.3](V1){} node[pos=0.6](V2){} node[pos=0.7](V6){} (O);
      \draw[antifermion] (I2) -- node[pos=0.3](V3){} node[pos=0.6](V4){} (O);
      \fill[counter, fill=white, draw] (O) circle (0.1);

      \draw[rgluon] (V1.center) -- (V4.center);
      \draw[rgluon] (V2.center) -- (V3.center);
      \draw[gluon=2.95pt] (V5.center) ..controls +(0.3,0.5) and +(0.3,0.5).. (V6.center);
    \end{tikzpicture}
      \caption{$(C_A-2C_F)(C_A-C_F)C_F$}
  \end{subfigure}

  \begin{subfigure}{.24\textwidth}
  \centering
  \begin{tikzpicture}[baseline]
    \coordinate (I1) at (-2,1.2);
    \coordinate (I2) at (-2,-1.2);
    \coordinate (O) at (0,0);

    \coordinate (V3) at (-1.2, 0.3);
    \coordinate (V4) at (-1.2, -0.3);

    \draw[fermion] (I1) -- node[pos=0.2](V5){} node[pos=0.4](V1){} node[pos=0.6](V6){} (O);
    \draw[antifermion] (I2) -- node[pos=0.4](V2){} (O);

    \fill[counter, fill=white, draw] (O) circle (0.1);

    \draw[rgluon] (V1.center) -- (V3);
    \draw[gluon] (V2.center) -- (V4);
    \draw[gluon=3.1pt] (V3) arc(90:-90:0.3);
    \draw[gluon=3.1pt] (V4) arc(-90:-270:0.3);

    \draw[gluon] (V5.center) ..controls +(0.3,0.5) and +(0.3,0.5).. (V6.center);
  \end{tikzpicture}
      \caption{$C_A(C_A-2C_F)C_F$}
  \end{subfigure}
  \begin{subfigure}{.24\textwidth}
  \centering
  \begin{tikzpicture}[baseline]
    \coordinate (I1) at (-2,1.2);
    \coordinate (I2) at (-2,-1.2);
    \coordinate (O) at (0,0);

    \draw[fermion] (I1) -- node[pos=0.3](V1){}  node[pos=0.65](V5){} (O);
    \draw[antifermion] (I2) -- node[pos=0.3](V2){} node[pos=0.65](V6){} (O);
    \fill[counter, fill=white, draw] (O) circle (0.1);

    \coordinate (V3) at (-1.4, 0.3);
    \coordinate (V4) at (-1.4, -0.3);

    \draw[rgluon] (V1.center) -- (V3);
    \draw[gluon] (V2.center) -- (V4);
    \draw[gluon=3.1pt] (V3) arc(90:-90:0.3);
    \draw[gluon=3.1pt] (V4) arc(-90:-270:0.3);

    \draw[rgluon] (V5.center) -- (V6.center);
  \end{tikzpicture}
      \caption{$C_AC_F^2$}
  \end{subfigure}
  \begin{subfigure}{.24\textwidth}
  \centering
    \begin{tikzpicture}[baseline]
      \coordinate (I1) at (-2,1.2);
      \coordinate (I2) at (-2,-1.2);
      \coordinate (O) at (0,0);
      \coordinate (V5) at (-1,0);

      \draw[fermion] (I1) -- node[pos=0.3](V1){} node[pos=0.7](V2){} (O);
      \draw[antifermion] (I2) -- node[pos=0.3](V3){} node[pos=0.7](V4){} (O);
      \fill[counter, fill=white, draw] (O) circle (0.1);

      \draw[gluon] (V5) -- (V1.center);
      \draw[gluon] (V5) -- (V2.center);
      \draw[gluon] (V5) -- (V3.center);
      \draw[gluon] (V5) -- (V4.center);
    \end{tikzpicture}
      \caption{$C_A^2C_F$}
  \end{subfigure}
  \begin{subfigure}{.24\textwidth}
  \centering
    \begin{tikzpicture}[baseline]
      \coordinate (I1) at (-2,1.2);
      \coordinate (I2) at (-2,-1.2);
      \coordinate (O) at (0,0);

      \draw[fermion] (I1) -- node[pos=0.3](V1){} (O);
      \draw[antifermion] (I2) -- node[pos=0.3](V2){} (O);
      \fill[counter, fill=white, draw] (O) circle (0.1);

      \coordinate (V3) at (-1.4,0.4);
      \coordinate (V4) at (-1.4,-0.4);

      \draw[rgluon] (V1.center) -- (V3);
      \draw[gluon] (V2.center) -- (V4);
      \draw[fermion] (V3) arc(90:0:0.4) node(V5){}; \draw (V5.center) arc(0:-90:0.4);
      \draw[fermion] (V4) arc(-90:-180:0.4) node(V6){} 
      ;
      \draw (V6.center) arc(-180:-270:0.4);
      \draw[rgluon] (V5.center) -- (V6.center);
    \end{tikzpicture}
      \caption{$(C_A-2C_F)C_F$}
  \end{subfigure}

  \begin{subfigure}{.24\textwidth}
  \centering
    \begin{tikzpicture}[baseline]
      \coordinate (I1) at (-2,1.2);
      \coordinate (I2) at (-2,-1.2);
      \coordinate (O) at (0,0);

      \draw[fermion] (I1) -- node[pos=0.2](V1){} node[pos=0.55](V2){} (O);
      \draw[antifermion] (I2) -- node[pos=0.2](V3){} node[pos=0.55](V4){} (O);
      \fill[counter, fill=white, draw] (O) circle (0.1);

      \coordinate (V5) at (-0.9,0.25) {};
      \coordinate (V6) at (-0.9,-0.25) {};

      \draw[rgluon] (V2.center) -- (V5);
      \draw[gluon] (V4.center) -- (V6);
      \draw[fermion] (V5) arc(90:-90:0.25) node[pos=0.5](fl){};
      \draw[fermion] (V6) arc(-90:-270:0.25);
      \draw[rgluon] (V1.center) -- (V3.center);
    \end{tikzpicture}
      \caption{$C_AC_F$}
  \end{subfigure}
  \begin{subfigure}{.24\textwidth}
  \centering
    \begin{tikzpicture}[baseline]
      \coordinate (I1) at (-2,1.2);
      \coordinate (I2) at (-2,-1.2);
      \coordinate (O) at (0,0);
      \coordinate (V4) at (-1.4,0.3);

      \draw[fermion] (I1) -- node[pos=0.2](V1){} node[pos=0.6](V2){} (O);
      \draw[antifermion] (I2) -- node[pos=0.3](V3){} (O);
      \fill[counter, fill=white, draw] (O) circle (0.1);

      \coordinate (V5) at (-1.4,0);
      \coordinate (V6) at (-1.4,-0.5);

      \draw[rgluon] (V1.center) -- (V4);
      \draw[gluon] (V2.center) -- (V4);
      \draw[gluon] (V5) -- (V4);
      \draw[fermion] (V5) arc(90:-90:0.25);
      \draw[fermion] (V6) arc(-90:-270:0.25) 
      ;
      \draw[rgluon] (V6) -- (V3.center);
    \end{tikzpicture}
      \caption{$C_F^2$}
  \end{subfigure}
  \begin{subfigure}{.24\textwidth}
  \centering
    \begin{tikzpicture}[baseline]
      \coordinate (I1) at (-2,1.2);
      \coordinate (I2) at (-2,-1.2);
      \coordinate (O) at (0,0);

      \draw[fermion] (I1) -- node[pos=0.15](V1){} (O);
      \draw[antifermion] (I2) -- node[pos=0.15](V2){} (O);
      \fill[counter, fill=white, draw] (O) circle (0.1);

      \coordinate (V3) at (-1.7,0.7);
      \coordinate (V4) at (-1.7,0.2);
      \coordinate (V5) at (-1.7,-0.2);
      \coordinate (V6) at (-1.7,-0.7);

      \draw[gluon] (V3.center) -- (V1.center);
      \draw[fermion] (V3) arc(90:-90:0.25);
      \draw[fermion] (V4) arc(-90:-270:0.25) 
      ;
      \draw[rgluon] (V4.center) -- (V5.center);
      \draw[fermion] (V5) arc(90:-90:0.25);
      \draw[fermion] (V6) arc(-90:-270:0.25) 
      ;
      \draw[rgluon] (V6.center) -- (V2.center);
    \end{tikzpicture}
      \caption{$C_F$}
  \end{subfigure}
  \begin{subfigure}{.24\textwidth}
  \centering
    \begin{tikzpicture}[baseline]
      \coordinate (I1) at (-1.4,0.8);
      \coordinate (I2) at (-1.4,-0.8);

      \coordinate (V1) at (-0.7,0.6);
      \coordinate (V2) at (-0.7,0);
      \coordinate (V3) at (-0.7,-0.6);
      \coordinate (V4) at (0,0.6);
      \coordinate (V5) at (0,0);
      \coordinate (V6) at (0,-0.6);

      \coordinate (O) at (1,0);

      \draw[fermion] (I1) -- (V1); \draw[fermion] (V1) -- (V2);
      \draw[fermion] (V2) -- (V3); \draw[fermion] (V3) -- (I2);

      \draw[gluon] (V1) -- (V4); \draw[gluon] (V2) -- (V5); \draw[gluon] (V3) -- (V6);

      \draw[fermion] (V4) -- (O); \draw[fermion] (O) -- (V6) 
      ;
      \draw[fermion] (V6) -- (V5); \draw[fermion] (V5) -- (V4);

      \fill[counter, fill=white, draw] (O) circle (0.1);
    \end{tikzpicture}
      \caption{$(4B_F\pm C_A)C_F$}
  \end{subfigure}
  \caption{Some representative diagrams for the direct channel. The color factor $B_F$ is defined as $\sum_{bc}d^{abc}d^{ebc} \equiv 4B_F\delta^{ae}$ and
  $B_F=(N_c^2-4)/(4N_c)$ for the $SU(N_c)$ group.
  %------------------
  \label{fig:feynmandiagrams-N3LO}}
  \end{figure}
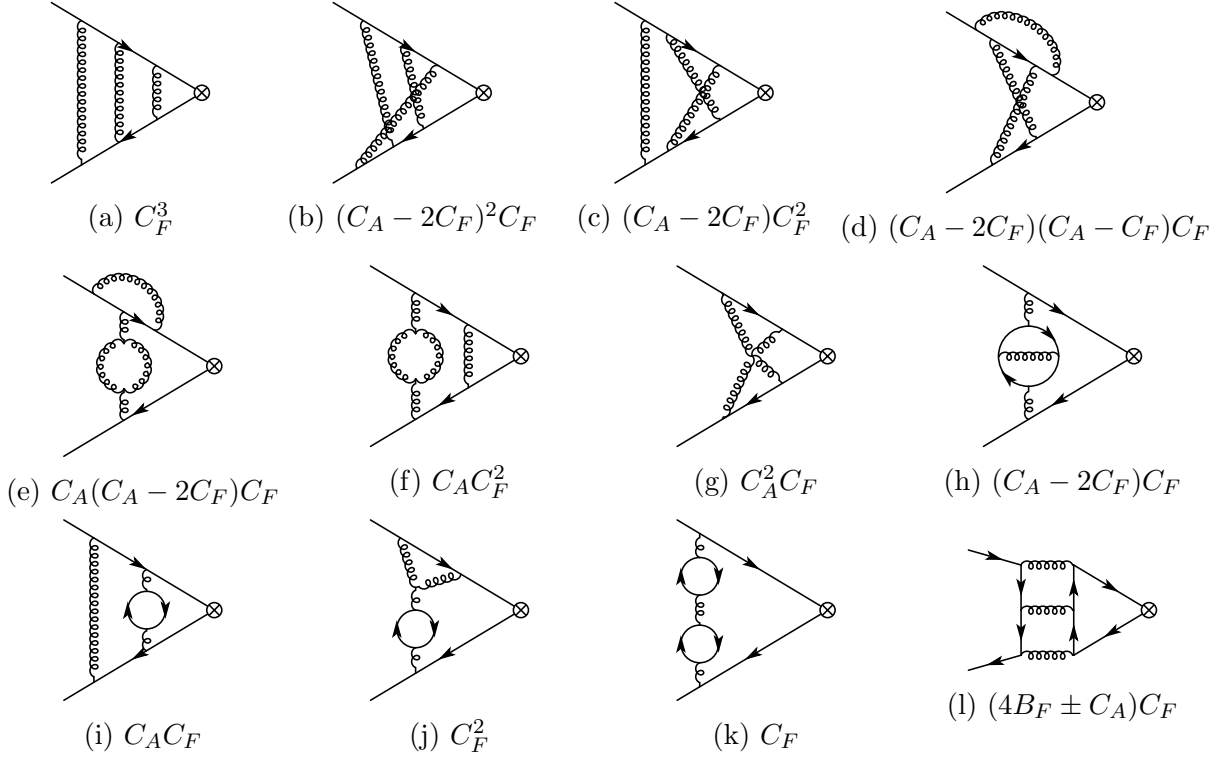

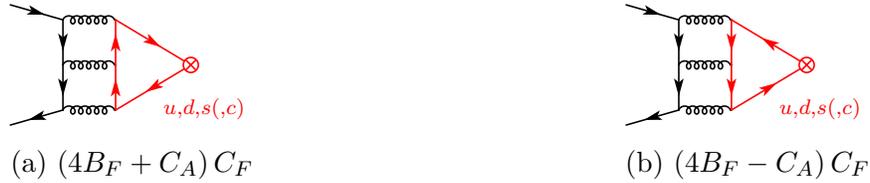
\begin{figure}
    \begin{subfigure}{.49\textwidth}
      \centering
      \begin{tikzpicture}[baseline]
      \coordinate (I1) at (-1.4,0.8);
      \coordinate (I2) at (-1.4,-0.8);

      \coordinate (V1) at (-0.7,0.6);
      \coordinate (V2) at (-0.7,0);
      \coordinate (V3) at (-0.7,-0.6);
      \coordinate (V4) at (0,0.6);
      \coordinate (V5) at (0,0);
      \coordinate (V6) at (0,-0.6);

      \coordinate (O) at (1,0);

      \draw[fermion] (I1) -- (V1); \draw[fermion] (V1) -- (V2);
      \draw[fermion] (V2) -- (V3); \draw[fermion] (V3) -- (I2);

      \draw[gluon] (V1) -- (V4); \draw[gluon] (V2) -- (V5); \draw[gluon] (V3) -- (V6);

      \draw[fermion, red] (V4) -- (O); \draw[fermion, red] (O) -- (V6) node[pos=0.5, below right]{$\scriptstyle{u,d,s(,c)}$};
      \draw[fermion, red] (V6) -- (V5); \draw[fermion, red] (V5) -- (V4);

      \fill[counter, fill=white, draw=red] (O) circle (0.1);
    \end{tikzpicture}
    \caption{$\left(4B_{F}+C_{A}\right)C_{F}$}
    \end{subfigure}
    \begin{subfigure}{.49\textwidth}
      \begin{tikzpicture}[baseline]
      \coordinate (I1) at (-1.4,0.8);
      \coordinate (I2) at (-1.4,-0.8);

      \coordinate (V1) at (-0.7,0.6);
      \coordinate (V2) at (-0.7,0);
      \coordinate (V3) at (-0.7,-0.6);
      \coordinate (V4) at (0,0.6);
      \coordinate (V5) at (0,0);
      \coordinate (V6) at (0,-0.6);

      \coordinate (O) at (1,0);

      \draw[fermion] (I1) -- (V1); \draw[fermion] (V1) -- (V2);
      \draw[fermion] (V2) -- (V3); \draw[fermion] (V3) -- (I2);

      \draw[gluon] (V1) -- (V4); \draw[gluon] (V2) -- (V5); \draw[gluon] (V3) -- (V6);

      \draw[antifermion, red] (V4) -- (O); \draw[antifermion, red] (O) -- (V6) node[pos=0.5, below right]{$\scriptstyle{u,d,s(,c)}$};
      \draw[antifermion, red] (V6) -- (V5); \draw[antifermion, red] (V5) -- (V4);

      \fill[counter, fill=white, draw=red] (O) circle (0.1);
    \end{tikzpicture}
    \caption{$\left(4B_{F}-C_{A}\right)C_{F}$}
    \end{subfigure}
    \caption{Some representative diagrams for the indirect channel. \label{fig:feynmandiagrams-indirect}}
  \end{figure}

The Feynman diagrams for $\Gamma$ are generated by the packages \texttt{QGraf}~\cite{Nogueira:1991ex} and \texttt{FeynArts}~\cite{Hahn:2000kx}. There are more than $300$
diagrams contributing to the vector current matching at $\text{N}^3$LO.   As illustrated in (\ref{eq:NRQCD factorization}), the amplitudes have been divided into two classes:
the direct one and indirect one. Some representative diagrams for each class are shown in FIG.~\ref{fig:feynmandiagrams-N3LO} and FIG.~\ref{fig:feynmandiagrams-indirect}, respectively. The latter contains a specific topology where a closed quark loop is linked with three gluons and the current. This topology is often denoted by the
``light-by-light'' or singlet diagrams in literature. We note that the singlet amplitudes in the direct channel has also been
considered in a very recent preprint that investigating the three-loop corrections to massive quark vector form factor~\cite{Boughezal:2022nof}.

 Since the sum of electric charge of three light quarks vanishes, \textit{i.e.}, $\sum_{l=u,d,s} e_l=0$,
 ${\mathcal{C}}_{\text{ind},l}^{(3)}$ can be safely ignored. Following the notation of literature~\cite{Marquard:2014pea}, we further decompose
 $\mathcal{C}^{(3)}_\text{dir}$ with respect to different color structures:
\begin{align}
  \mathcal{C}_{\text{dir}}^{(3)}= &C_F \big[ C_F^2 \mathcal{C}_{FFF} + C_F C_A \mathcal{C}_{FFA} + C_A^2 \mathcal{C}_{FAA}
  \nn\\
  &+ T_F n_L\left(
  C_F \mathcal{C}_{FFL} + C_A \mathcal{C}_{FAL} + T_F n_H \mathcal{C}_{FHL} + T_F n_M \mathcal{C}_{FML} + T_F n_L \mathcal{C}_{FLL}
  \right)
  \nn\\
  &+ T_F n_H\left(
  C_F \mathcal{C}_{FFH} + C_A \mathcal{C}_{FAH} + T_F n_H \mathcal{C}_{FHH}+T_F n_M \mathcal{C}_{FHM}
  +B_F\mathcal{C}_{BFH}\right)\nn\\
  &+ T_F n_M\left(
  C_F \mathcal{C}_{FFM} + C_A \mathcal{C}_{FAM} + T_F n_M \mathcal{C}_{FMM}
  \right) \big].
\end{align}

In extracting the SDCs, we have employed the covariant projector technique to project the amplitude with free $Q\bar{Q}$ pair onto the desired quantum number ${}3S_1^{(1)}$.
The packages {\tt FeynCalc}/{\tt FormLink}~\cite{Mertig:1990an,Feng:2012tk} are then utilized to deal with the trace over Dirac and $SU(N_c)$ color matrices. There are about {$300$} master integrals (MIs) for the amplitudes after the integration-by-parts (IBP) reduction with the aid of {\tt Apart}~\cite{Feng:2012iq} and {\tt FIRE}~\cite{Smirnov:2014hma}. The `light-by-light" amplitudes have imaginary parts, making the numerical evaluation of these MIs a painful
challenge for traditional numerical methods such as sector decomposition~\cite{Hepp:1966eg}.
We instead turn to the newly developed \texttt{AMF} package based on numerical differential equation technique,
which can compute the multi-loop MIs to a very high precision in a very effective way~\cite{Liu:2017jxz,Liu:2020kpc,Liu:2022chg}.

\subsection{Reconstructing the NRQCD Renormalization Constant}

To remove the UV divergences, we incorporate the on-shell field and mass renormalization by taking the order-$\alpha_s^3$ expressions of $Z_2$ and $Z_m$ from \cite{Broadhurst:1991fy, Melnikov:2000zc,Marquard:2007uj}.
The strong coupling constant is renormalized to two-loop order under $\overline{\rm MS}$ scheme.
After the renormalization procedure, the amplitude still
contains uncancelled IR poles. These IR divergences appearing in the hard region in QCD amplitude is an indicator that the NRQCD current requires additional renormalization,
which should be cancelled by the UV divergences in $\tilde{Z}_v$.
Therefore, one can calculate the coefficients of the single poles with different values of the mass ratio $x$,
then reconstruct $\gamma_0$ through numerical fitting recipe:
\begin{align}
  \gamma_0=\dfrac{C_Fm_Q^2}{60m_M^2}+\left(\dfrac{C_F}{18}+\dfrac{C_A}{12}\right)\ln\dfrac{m_M^2}{m_Q^2}.
\end{align}
This completes our knowledge about the renormalization constant $\tilde{Z}_v$. We are then able to deduce the complete expression for
the anomalous dimension up to three-loop:
\begin{align}
  \gamma_v^{(2)}=&-3 \uppi^2 C_F \left(\dfrac{1}{18}C_F+\dfrac{1}{12}C_A\right),\nn \\
  \gamma_v^{(3)}\left(\mu_\Lambda\right)=&-3 \uppi^2 C_F \bigg\{\left(\dfrac{43}{144}-\dfrac{1}{2}\ln 2\right)C_F^2+\left(\dfrac{113}{324}+\dfrac{1}{4}\ln 2\right)C_F C_A+\left(\dfrac{2}{27}+\dfrac{1}{4}\ln 2\right)C_A^2 \nn \\
  & +T_F n_L \left(-\dfrac{25}{324}C_F-\dfrac{37}{432}C_A\right)+\dfrac{1}{60}T_F n_H C_F\nn\\
  &+T_F n_M\left[\dfrac{1}{60x^2}+\left(\dfrac{1}{18} C_F + \dfrac{1}{12} C_A \right)\ln x^2\right]+\Lmu\bigg[\dfrac{5}{48}C_F^2+\dfrac{5}{32}C_F C_A\nn\\
  &+\dfrac{1}{24}C_A^2-T_F n_M\left(\dfrac{1}{12}C_F+\dfrac{1}{8}C_A\right)\bigg]\bigg\}.
\end{align}

The $n_M$-independent parts of SDCs are independent of the mass ratio $x$. These SDCs, except for $\mathcal{C}_{BFH}$ and ${\mathcal{C}}_{\text{ind},l}^{(3)}$, have been evaluated in \cite{Marquard:2006qi,Marquard:2009bj,Marquard:2014pea,Egner:2022jot}. $\mathcal{C}_{FHL}$, $\mathcal{C}_{FLL}$ and $\mathcal{C}_{FHH}$ are calculated analytically, while others are known numerically. Our numerical results confirm the known results, but with much higher precision:
\begin{align}
  \mathcal{C}_{FFF} =&{36.49486245880592537633476189872792031664181},\nn\\
  \mathcal{C}_{FFA} =&{-188.07784165988071390579994023278476450389105},\nn\\
  \mathcal{C}_{FAA} =&{-97.734973269918386342345245004574098439887181},\nn\\
  \mathcal{C}_{FFL} =&{46.691692905515132467558267641260536017779126774},\nn\\
  \mathcal{C}_{FAL} =&{39.6237185545244190773420474220534775186981204767},\nn\\
  \mathcal{C}_{FHL} =&{-0.270250439156502171732138691397778647923997721},
  \nn\\
  \mathcal{C}_{FLL} =& {-2.46833645448237411637054187652486189658968386},
  \nn\\
  \mathcal{C}_{FFH} =&{-0.8435622911595001453055093736419593585798252},
  \nn\\
  \mathcal{C}_{FAH} =&{-0.1024741614929317408574835971993802120163106},
  \nn\\
  \mathcal{C}_{FHH} =&{0.05123960751198372493493118588999641369844635617},\nn\\
  \mathcal{C}_{BFH}=&{2.1155782679809064984368222219139443700443356}\nn\\
  &{+\mathrm{i}\,0.494212710700672040241218108020160381155220487)},\nn\\
  {\mathcal{C}}_{\text{ind},l}^{(3)}=&T_FB_FC_F(-0.945532642977386 +{\mathrm{i}\,1.28500237447426)}.
\end{align}

\begin{figure}
  \includegraphics[width=0.32\textwidth]{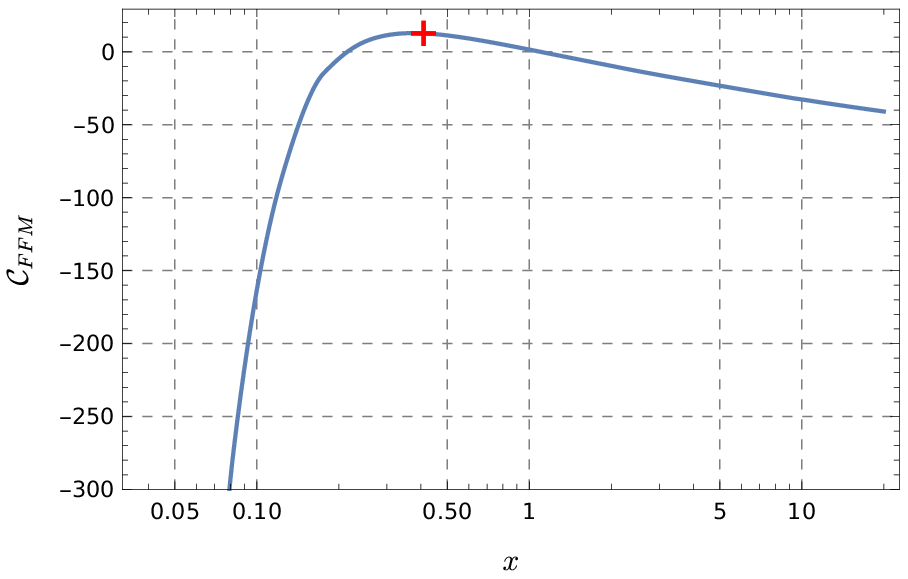}
  \includegraphics[width=0.32\textwidth]{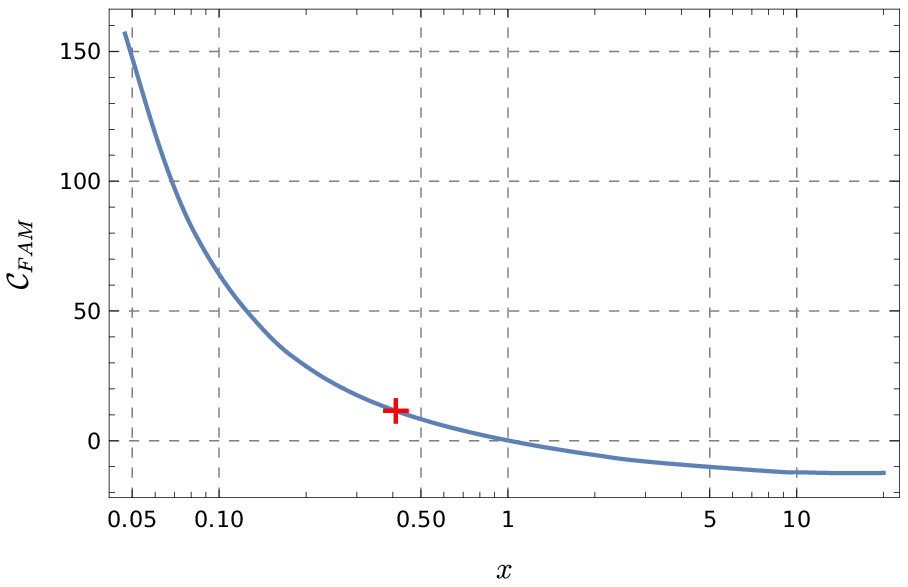}
  \includegraphics[width=0.32\textwidth]{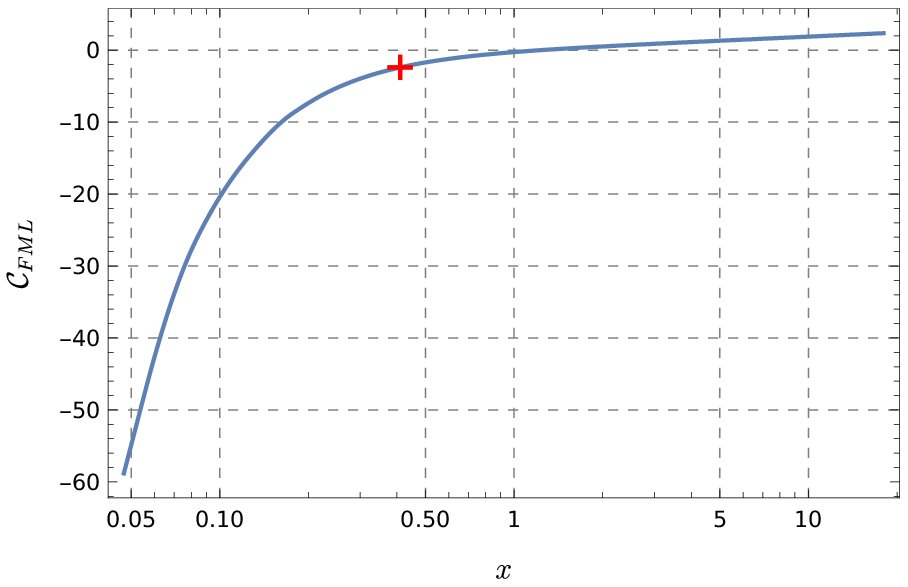}
  \includegraphics[width=0.32\textwidth]{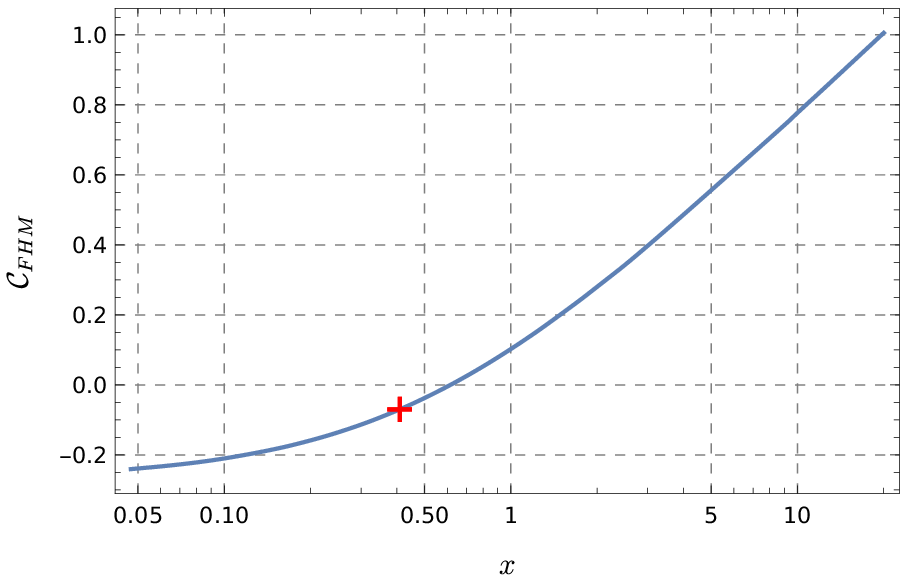}
  \includegraphics[width=0.32\textwidth]{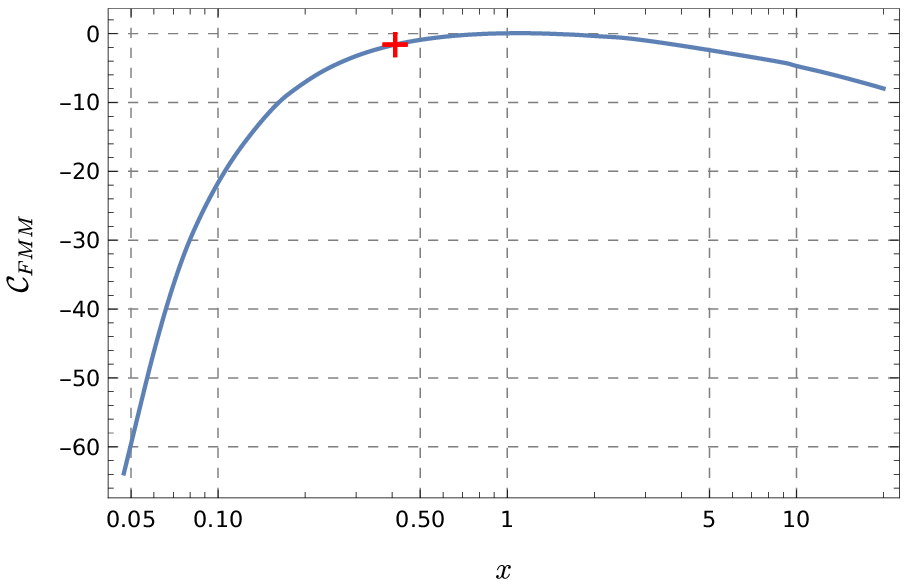}
  \includegraphics[width=0.32\textwidth]{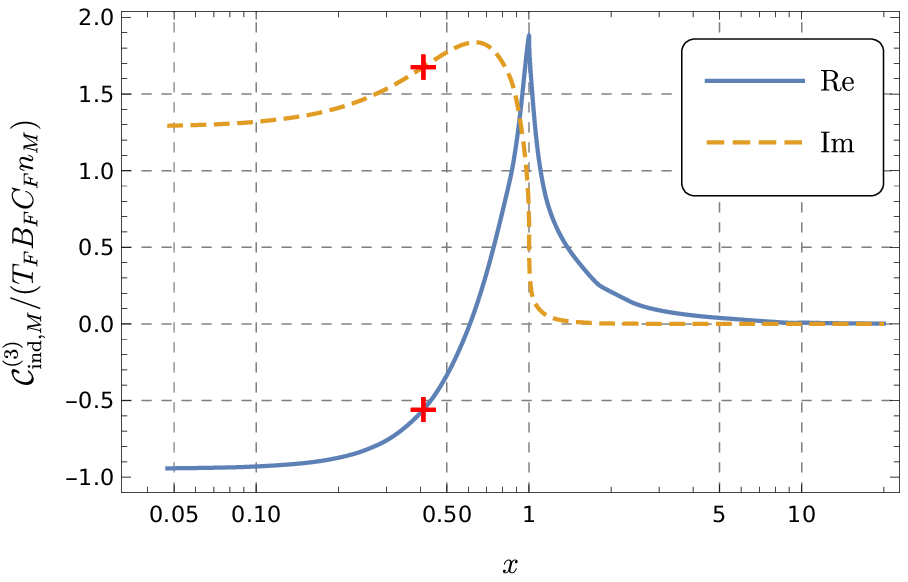}
  \caption{$n_M$-dependent terms of the SDCs as function of $x$. The crosses label the SDCs for $\Upsilon$ leptonic decay with charm quark mass $2.04$ GeV and bottom quark
  mass $4.98$ GeV. with physical mass ratio. \label{fig:nm terms}}
\end{figure}

Since the $n_M$-dependent terms in the SDCs vary with $x$, they are complicated functions of $x$ instead of constants. We have
numerically evaluate these terms with many different values of $x$ and show their profiles in FIG.~\ref{fig:nm terms}.

For reader's convenience, we take the three-loop pole mass to be $m_c=2.04$ GeV and $m_b=4.98$ GeV, and define $x_\text{phy}$
as the corresponding ratio.
We explicitly present the numerical SDCs for $\Upsilon$ decay with $x_\text{phy}=2.04/4.98$ as follows:
\begin{align} \mathcal{C}_{FFM}\left(x_\text{phy}\right) =&12.59634873592650205755866578463730117088523164,
  \nn\\
  \mathcal{C}_{FAM}\left(x_\text{phy}\right) =&11.485487450937888599629526907235667539099062,
  \nn\\
  \mathcal{C}_{FML}\left(x_\text{phy}\right) =&-2.40343737830471435766973034309185481125317985382,
  \nn\\
  \mathcal{C}_{FHM}\left(x_\text{phy}\right) =&-0.0697639123255824045909089712436335822048386627,
  \nn\\
  \mathcal{C}_{FMM} \left(x_\text{phy}\right)=& -1.59441395438374277579099808902128228543581741117,\nn\\
  {\mathcal{C}}_{\text{ind},M}^{(3)}\left(x_\text{phy}\right)=&T_FB_FC_Fn_M(-0.5600172596682562131259719921291458956964640\nn\\
  &{\mathrm{i}\,1.675430934114461038053844582307006798974491776)}.
\end{align}

We enumerate the SDCs at various perturbative order with $\mu_R=m_b$ and $\mu_\Lambda=1.5\,\mathrm{GeV}$. For the sake of clarity,
we retain the explicit dependence on $n_L, n_H, n_M$ and separate out the ``light-by-light'' contributions at $\text{N}^3$LO:
\begin{align}
  \mathcal{C}&\left({\mu_R=m_b,\mu_\Lambda=1.5\,\mathrm{GeV},x_\text{phy}}\right)\approx 1
  -{\dfrac{8}{3}} \dfrac{\alpha_s(m_b)}{\uppi}{-11.168625031474922}\left(\dfrac{\alpha_s(m_b)}{\uppi}\right)^2\nn\\
  &+ \left(\dfrac{\alpha_s(m_b)}{\uppi}\right)^3
  \big[{-1702.6786280810694238919125522+91.423820122192320694764194n_L}\nn\\
  &{+0.097945436227233777101707796n_H+19.1788305723128699826853781n_M}\nn\\
  &{-0.82277881816079137212351396 n_L^2+0.0170798691706612416449770620n_H^2}\nn\\
  &{-0.53147131812791425859699936 n_M^2-0.80114579276823811922324345n_Ln_M}\nn\\
  &{-0.09008347971883405724404623n_L n_H -0.0232546374418608015303029904n_Mn_H}\nn\\
  &+(0.5876606299946962495657840 + {\mathrm{i}\,0.1372813085279644556225606)_{\text{ind},H}}\nn\\
  &+(0.3111206998156978961810956-{\mathrm{i}\,0.9307949633969227989188025)_{\text{ind},M}}
    \big].
\end{align}

\section{Phenomenology \label{sec:phenomenology}}

With the complete three-loop SDCs in hand, we are able to present a finest NRQCD predictions to the leptonic width of vector quarkonium:
\begin{align}
  \Gamma(V\rightarrow l^+l^-)={\dfrac{4\uppi\alpha^2 }{3M_V}|f_V|^2}
  =&\dfrac{8\uppi\alpha^2e_Q^2}{3M_V^2}\left|\mathcal{C}\left(\dfrac{\mu_R}{m_Q},\dfrac{\mu_\Lambda}{m_Q},x\right)\right|^2
  \left|\left\langle 0 \vert \chi^\dagger\boldsymbol{\upsigma}\cdot\boldsymbol{\upepsilon}\psi(\mu_\Lambda) \vert V(\boldsymbol\upepsilon) \right\rangle\right|^2\nn\\
  =&\dfrac{8\uppi\alpha^2e_Q^2}{3M_V^2}
  \left|\mathcal{C}_\text{dir}+\sum_{f\neq Q}\mathcal{C}_\text{ind,f}\dfrac{e_f}{e_Q}\right|^2
  \left|\left\langle 0 \vert \chi^\dagger\boldsymbol{\upsigma}\cdot\boldsymbol{\upepsilon}\psi(\mu_\Lambda) \vert V(\boldsymbol\upepsilon) \right\rangle\right|^2.
\end{align}

The nonperturbative NRQCD matrix element can be estimated in variolous theoretical approach. It has been investigated in lattice NRQCD~\cite{Choe:2003wx,Gray:2005ur}.
In practice, it is often estimated in potential quark models and expressed in terms of $R_V(0)$, the radial Schr\"odinger function at the origin:
\begin{align}
  \left\langle 0 \vert \chi^\dagger\boldsymbol{\upsigma}\cdot\boldsymbol{\upepsilon}\psi(\mu_\Lambda) \vert V(\boldsymbol\upepsilon) \right\rangle=\sqrt{\dfrac{N_c}{2\uppi}}R_V(0)
\end{align}

The exact value of $R_V(0)$ varies with the different potential models, which constitutes a major source of theoretical uncertainties.
For reader's convenience, in TABLE~\ref{table:LDME} we tabulate its value estimated in different potential models~\footnote{The Bohr result is evaluated via the formula $R_{V}^2(0)=32m_Q^3\alpha_s^3(\mu_R)/27$, where $m_b^\text{B}=5.059$ and $m_c^\text{B}=1.65\,\mathrm{GeV}$. $\mu_R$ ranges from $1.5\,\mathrm{GeV}$ to $2m_b$ and $1\,\mathrm{GeV}$ to $2m_c$ for $\Upsilon$ and $\jpsi$, respectively. The central values correspond to $\mu_R=3.5$ and $2\,\mathrm{GeV}$.}.

\renewcommand\arraystretch{1.2}

\begin{table}[h]
  \begin{tabular}{|c|c|c|c|c|c|c|}
    \hline
    Potential Model
    & \tabincell{c}{Cornell\\~\cite{Eichten:1995ch}}
    & \tabincell{c}{Lattice\\~\cite{Choe:2003wx,Gray:2005ur}}
    & \tabincell{c}{B-T\\~\cite{Eichten:1995ch}}
    &\tabincell{c}{Coul. $+$ power\\~\cite{Rai:2008sc}}
    &\tabincell{c}{Bohr\\~\cite{Beneke:2014qea,Egner:2021lxd}}
    &\tabincell{c}{pNRQCD\\~\cite{Chung:2020zqc}}
    \\
    \hline
    $|R_{\Upsilon}(0)|^2\,(\mathrm{GeV}^{3})$
    &$14.05$
    &$5.05613$
    &$6.477$
    &$3.909\sim 9.181$
    &{$2.1094^{+4.5371}_{-1.2913}$}
    & $3.092$\\
    \hline
    $|R_{\jpsi}(0)|^2\,(\mathrm{GeV}^{3})$
    &$1.454$
    &$1.1184$
    &$0.810$
    &$0.610\sim 1.850$
    &{$0.1423^{+0.5007}_{-0.0860}$}
    & $0.421$\\
    \hline
    Potential Model
    & \tabincell{c}{Screened\\~\cite{Azhothkaran:2020ipl}}
    & \tabincell{c}{Power Law\\~\cite{Eichten:1995ch}}
    & \tabincell{c}{Log\\~\cite{Eichten:1995ch}}
    & \tabincell{c}{Modified NR\\~\cite{Akbar:2015evy,Akbar:2011jd}}
    &  \tabincell{c}{Semi-relativistic\\~\cite{Radford:2007vd}}
    &Coul.\\
    \hline
    $|R_{\Upsilon}(0)|^2\,(\mathrm{GeV}^{3})$
    &$8.72$
    &$4.591$
    &$4.916$
    &$11.4185$
    & $6.143$
    &$4.221$\\
    \hline
    $|R_{\jpsi}(0)|^2\,(\mathrm{GeV}^{3})$
    &$1.19$
    &$0.999$
    &$0.815$
    &$1.9767$
    & $0.478$
    &$0.303$\\
    \hline
  \end{tabular}
  \caption{The values of $R_V(0)$ estimated in different theoretical approaches. \label{table:LDME}
  Bowdin \textit{et al.} \cite{Bodwin:2007fz} also determined the value of LDMEs, and the results are not included here
  since their computation is based on the measured leptonic width as input.}
\end{table}

In phenomenological study, we adopt the central value of $R_V(0)$  from the Buchm\"uller-Tye(BT) model~\cite{Eichten:1995ch}. To estimate the error caused by the uncertainty of
the wave function at the origin, one may simply rescale the phenomenological predictions by a factor $\kappa$ to float the LDMEs. 
According to TABLE~\ref{table:LDME}, one finds that $\kappa$ ranges from $0.1$ to $2.2$ for $\Upsilon$, and $0.07$ to $2.3$ for $\jpsi$. 

To make concrete phenomenological predictions, we take the following values for various input parameters:
\begin{align}
  &{m_b=4.98\,\mathrm{GeV}},\quad{\alpha(2m_b)=\dfrac{1}{132.5}},\quad\mu_\Lambda(\Upsilon)=1.5\,\mathrm{GeV},
  \nn\\
  &{m_c=2.04\,\mathrm{GeV}},\quad{\alpha(2m_c)=\dfrac{1}{133.7}},\quad\mu_\Lambda(\jpsi)=1.0\,\mathrm{GeV}.
\end{align}
We compute the values bottom and charm quark pole mass using the three-loop formula by taking the precisely known $\overline{\text{MS}}$ masses as input. 
The running QED/QCD coupling constants $\alpha$/$\alpha_s$ are evaluated using the packages 
\texttt{alphaQED}~\cite{Jegerlehner:2011mw} and \texttt{RunDec}~\cite{Herren:2017osy}, respectively.

In TABLE~\ref{table:decay width}, we present the NRQCD predictions at the various levels of perturbative accuracy for the leptonic width of $\Upsilon$ and $\jpsi$, juxtaposed with
the precise experimental results. For $\jpsi$ decay, the bottom contributions are not included, \textit{i.e.}, $n_M=0$. 
The finite charm mass effect appears to be noticeable for the $\Upsilon$ leptonic decay, while the indirect contributions are completely insignificant. 
It seems that after including the $\text{N}^3$LO corrections, the finest NRQCD predictions for $\Upsilon$ and $\jpsi$ are much smaller than the experimental data.

\begin{figure}
  \includegraphics[width=0.45\textwidth]{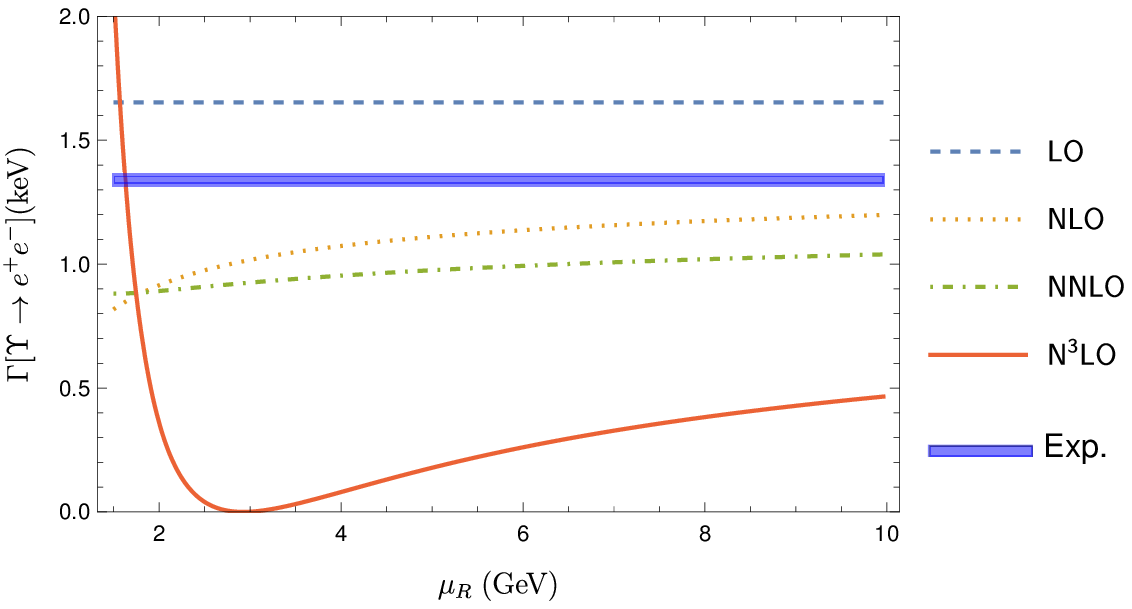}
  \includegraphics[width=0.45\textwidth]{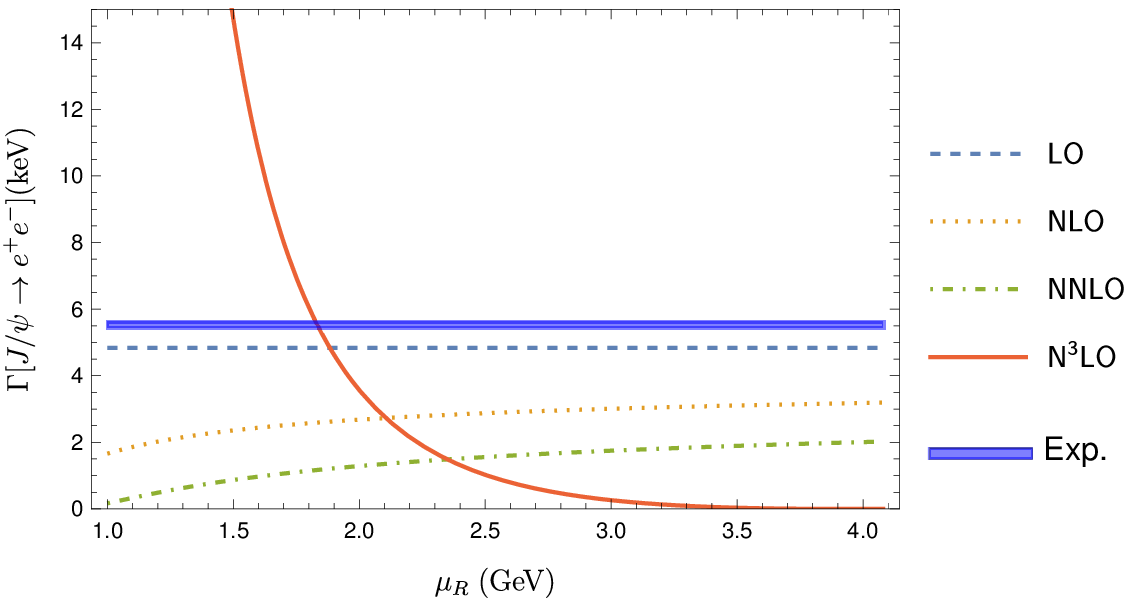}
  \caption{NRQCD predictions for leptonic width of $\Upsilon$ and $\jpsi$ as a
  function of renormalization scale $\mu_R$ at various level of perturbative expansion.\label{fig:muR-dependence}}
\end{figure}

\begin{figure}
  \includegraphics[width=0.9\textwidth]{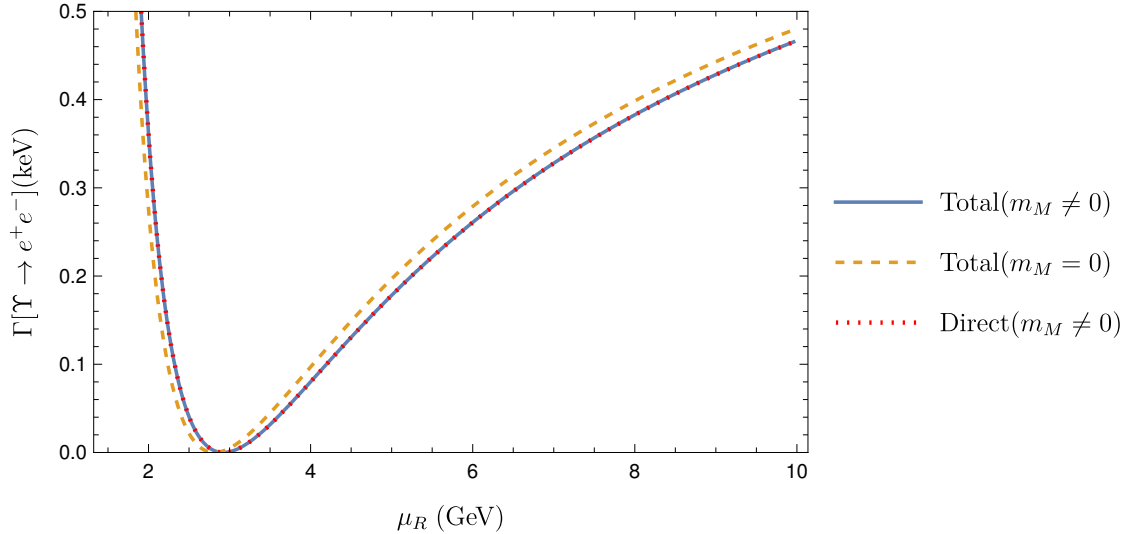}
  \caption{The $\text{N}^3$LO leptonic width of $\Upsilon$ as a
  function of renormalization scale $\mu_R$. One can see that the finite charm mass brings in a noticeable effect, but 
  the indirect channel has a completely negligible impact. \label{fig:muR-dependence-masseffect}}
\end{figure}

In FIG.~\ref{fig:muR-dependence} we plot our NRQCD predictions for leptonic width as a function of $\mu_R$ at different levels of perturbative order. At $\text{N}^3$LO, the renormalization scale dependence becomes much worse than the lower orders. The reason is that the SDCs $\mathcal{C}_{FFA}$ and $\mathcal{C}_{FAA}$ are very large. In FIG.~\ref{fig:muR-dependence-masseffect}, we plot the $\text{N}^3$LO decay rate of $\Upsilon$ for both massive and massless charm cases.

\begin{table}
  \begin{tabular}{|c|c|c|c|c|c|c|c|}
    \hline
    \multirow{3}{*}{\diagbox[height=3.6\line,width=50pt]{V}{%
    {$\Gamma(\mathrm{keV})$}}}&\multirow{3}{*}{LO}&\multirow{3}{*}{NLO}&\multirow{3}{*}{NNLO}&\multicolumn{3}{c|}{$\text{N}^3$LO}&\multirow{3}{*}{PDG}\\\cline{5-7}
    &&&&\tabincell{c}{Direct\\($m_M=0$)}&\tabincell{c}{Direct\\($m_M\neq0$)}&Total&\\
    \hline
    $\Upsilon$
    &${1.6529}$
    &${{1.1095}_{-0.2922}^{+0.0888}}$
    &${{0.9750}_{-0.0942}^{+0.0642}}$
    &${0.1948_{-0.1948}^{+1.5900}}$
    &${{0.1763}_{-0.1763}^{+1.9577}}$
    &${{0.1764}_{-0.1764}^{+1.9560}}$
    &$1.340\pm0.018$\\
    \hline
    $\jpsi$
    &${4.8392}$
    &${2.6999_{-1.0391}^{+0.4925}}$
    &${1.3138_{-1.1444}^{+0.7094}}$
    &\multicolumn{3}{c|}{${3.2219_{-3.2219}^{+123.4838}}$}&$5.53\pm0.10$\\
    \hline
  \end{tabular}
  \caption{Leptonic widths of $\jpsi$ and $\Upsilon$. The central values of predictions are obtained by setting $\mu_R=m_Q$\label{table:decay width}, while the errors are estimated by varying $\mu_R$ from $\mu_\Lambda$ to $2m_Q$.}
\end{table}


\section{Summary\label{sec:summary}}

A complete $\text{N}^3$LO analysis of the $\Upsilon/\jpsi$ leptonic width is presented within the NRQCD factorization framework. The SDCs for both direct and indirect channels are calculated numerically with exquisitely high precision. We also consider the finite charm effect for $\Upsilon$ decay, also find a novel contribution to the anomalous dimension 
of NRQCD vector current for $\Upsilon$ decay that arises from keeping charm quark massive. 

Adopting the values of the wave functions at the origin for $J/\psi$ and $\Upsilon$ from popular potential models, 
we evaluate the leptonic width of $\Upsilon$ and $J/\psi$ at $\text{N}^3$LO in $\alpha_s$. Unfortunately, the predictions exhibits a rather strong dependence on renormalization scale. For the natural range of $\mu_R$, there appears to exist an alarming discrepancy between the finest NRQCD predictions and the experimental data.
How to resolve this discrepancy definitely deserves further investigation. 

\begin{acknowledgments}
  %-----------------------
  {\noindent\it Acknowledgment.}
  %-----------------------
  The work of F. F. is supported by the National
Natural Science Foundation of China under Grant No. 11875318, No. 11505285, and by
the Yue Qi Young Scholar Project in CUMTB.
  The work of Y.~J., Z.~M., J.~P and J.-Y.~Z. is supported in part by
  the National Natural Science Foundation of China under Grants No. 11925506, 11875263,
  No. 11621131001 (CRC110 by DFG and NSFC).
  The work of W.-L. S. is supported by the National Natural Science Foundation of China
under Grants No. 11975187 and the Natural Science Foundation of ChongQing under
Grant No. cstc2019jcyj-msxmX0479.
  \end{acknowledgments}

\appendix
\section{$5$-Flavor}
\begin{figure}[h]
  \includegraphics[]{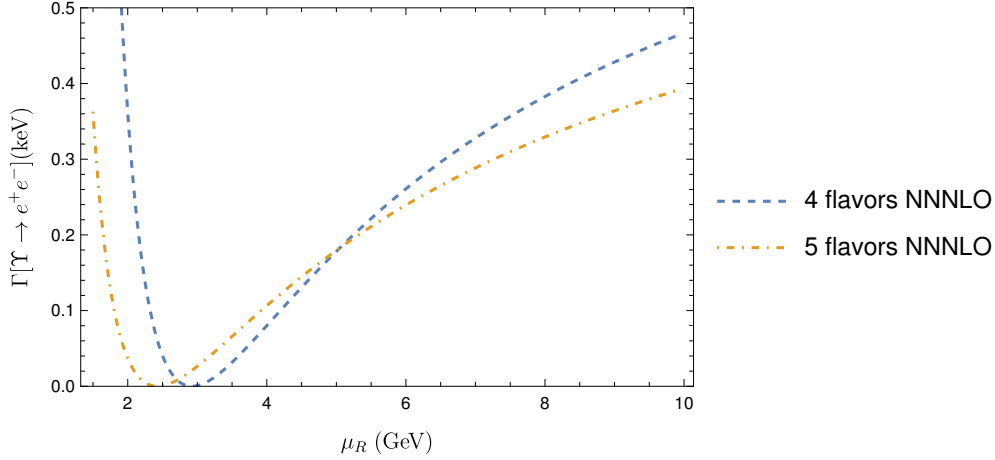}
  \caption{$\text{N}^3$LO NRQCD predictions for leptonic decay of $\Upsilon$ and $\jpsi$ as a
  function of renormalization scale $\mu_R$ with different active flavors.\label{fig:muR-dependence-5f}}
\end{figure}

Our main results are expressed as a power series w.r.t. $\alpha_s^{(n_L+n_M)}$ since the heavy quark is decoupled.
In practice, we first calculate $\Gamma$ with $n_L+n_M+n_H$ active flavors in QCD and then apply the following formula to decouple the heavy quark from the running of $\alpha_s$.
\begin{align}
    \dfrac{\alpha_{s}^{\left(n_{L}+n_{M}+n_{H}\right)}\left(\mu_{R}\right)}{\pi}
    =&\dfrac{\alpha_{s}^{\left(n_{L}+n_{M}\right)}\left(\mu_{R}\right)}{\pi}
    \nn\\
 &+\left(\dfrac{\alpha_{s}^{\left(n_{L}+n_{M}\right)}\left(\mu_{R}\right)}{\pi}\right)^{2}
 T_{F}n_{H}
 \left[\dfrac{1}{3}\Lmu+\left(\dfrac{1}{6}\ln^{2}\dfrac{\mu_\Lambda}{m_Q}+\dfrac{1}{36}\pi^{2}\right)\epsilon+\mathcal{O}\left(\epsilon^{2}\right)\right]
 \nn\\
 &+\left(\dfrac{\alpha_{s}^{\left(n_{L}+n_{M}\right)}\left(\mu_{R}\right)}{\pi}\right)^{3}
 T_{F}n_{H}
 \Bigg[\left(\dfrac{1}{4}\Lmu+\dfrac{15}{16}\right)C_{F}+\left(\dfrac{5}{12}\Lmu-\dfrac{2}{9}\right)C_{A}
 \nn\\
 &+\dfrac{1}{9}T_{F}n_{H}\ln^{2}\dfrac{\mu_\Lambda}{m_Q}+\mathcal{O}\left(\epsilon\right)\Bigg]+{\cal O}\left(\alpha_{s}^{4}\right)
\end{align}
Here we also present the results with $n_L+n_M+n_H$ active flavors. To deal with the remaining IR divergence, one should adopt the corresponding renormalization constant with with $n_L+n_M+n_H$ active flavors. For simplicity, we omit the superscript $(n_L+n_M+n_H)$ of $\alpha_s$ in the rest of paper.
\begin{align}
  \tilde{Z}_v=&1 + \left(\dfrac{\alpha_s(\mu_\Lambda)}{\uppi}\right)^2
  \dfrac{C_F\uppi^2}{\epsilon}\left(\dfrac{1}{12} C_F + \dfrac{1}{8} C_A \right)
    + \left(\dfrac{\alpha_s(\mu_\Lambda)}{\uppi}\right)^3  C_F\uppi^2\nn\\
    &\times \Bigg\{ C_F^2 \left[\dfrac{5}{144\epsilon^2} + \left(
    \dfrac{43}{144} - \dfrac{1}{2} \ln2 + \dfrac{5}{48} \Lmu
    \right)\dfrac{1}{\epsilon} \right]\nn\\
 & +C_FC_A \left[\dfrac{1}{864\epsilon^2} + \left( \dfrac{113}{324} +
    \dfrac{1}{4} \ln2 + \dfrac{5}{32} \Lmu \right) \dfrac{1}{\epsilon}
    \right]\nn\\
  &+C_A^2 \left[-\dfrac{1}{16\epsilon^2} + \left( \dfrac{2}{27} +
    \dfrac{1}{4} \ln2 + \dfrac{1}{24} \Lmu \right) \dfrac{1}{\epsilon}
    \right]\nn\\
 & + T_F n_L \left[ C_F\left(
      \dfrac{1}{54\epsilon^2}
      -\dfrac{25}{324\epsilon}
    \right)
    + C_A \left(
      \dfrac{1}{36\epsilon^2}
      - \dfrac{37}{432\epsilon} \right)
     \right]\nn\\
     &+ T_F n_H \left[\dfrac{C_F }{60\epsilon}-\left(\dfrac{1}{18} C_F + \dfrac{1}{12} C_A \right)\Lmu \dfrac{1}{\epsilon}\right]\nn\\
     &+T_Fn_M \left[\dfrac{1}{\epsilon}\gamma_0-\left(\dfrac{1}{18} C_F + \dfrac{1}{12} C_A \right)\Lmu\dfrac{1}{\epsilon}\right]
  \Bigg\} + {\cal O}(\alpha_s^4)\,,
\end{align}
The SDCs are the same as the $n_L+n_H$-active-flavor case except for:
\begin{align}
  \mathcal{C}_{FFH} =& {1.482405722886807434492825499916861}
  \nn\\
  \mathcal{C}_{FAH} =&{0.129533415132085184395574268694405}
\end{align}
We plot the $\text{N}^3$LO NRQCD predictions for the decay rate as a function of
$\mu_R$ with different active flavor numbers in FIG.~\ref{fig:muR-dependence-5f}. The $n_L+n_M+n_H$-active-flavor has a milder dependence on the renormalization scale.


\begin{thebibliography}{99}

%\cite{Hatton:2020qhk}
\bibitem{Hatton:2020qhk}
D.~Hatton \textit{et al.} [HPQCD],
%``Charmonium properties from lattice $QCD$+QED : Hyperfine splitting, $J/\psi$ leptonic width, charm quark mass, and $a^c_\mu$,''
Phys. Rev. D \textbf{102}, no.5, 054511 (2020)
doi:10.1103/PhysRevD.102.054511
[arXiv:2005.01845 [hep-lat]].
%37 citations counted in INSPIRE as of 13 Jul 2022

%\cite{Hatton:2021dvg}
\bibitem{Hatton:2021dvg}
D.~Hatton, C.~T.~H.~Davies, J.~Koponen, G.~P.~Lepage and A.~T.~Lytle,
%``Bottomonium precision tests from full lattice QCD: Hyperfine splitting, \Upsilon{} leptonic width, and b quark contribution to $e^+e^- \rightarrow$ hadrons,''
Phys. Rev. D \textbf{103}, no.5, 054512 (2021)
doi:10.1103/PhysRevD.103.054512
[arXiv:2101.08103 [hep-lat]].
%10 citations counted in INSPIRE as of 13 Jul 2022

%\cite{Caswell:1985ui}
\bibitem{Caswell:1985ui}
W.~E.~Caswell and G.~P.~Lepage,
%``Effective Lagrangians for Bound State Problems in QED, QCD, and Other Field Theories,''
Phys. Lett. B \textbf{167}, 437-442 (1986)
doi:10.1016/0370-2693(86)91297-9
%1213 citations counted in INSPIRE as of 06 Jul 2022

%\cite{Bodwin:1994jh}
\bibitem{Bodwin:1994jh}
G.~T.~Bodwin, E.~Braaten and G.~P.~Lepage,
%``Rigorous QCD analysis of inclusive annihilation and production of heavy quarkonium,''
Phys. Rev. D \textbf{51}, 1125-1171 (1995)
[erratum: Phys. Rev. D \textbf{55}, 5853 (1997)]
doi:10.1103/PhysRevD.55.5853
[arXiv:hep-ph/9407339 [hep-ph]].
%2683 citations counted in INSPIRE as of 06 Jul 2022

%\cite{VanRoyen:1967nq}
\bibitem{VanRoyen:1967nq}
R.~Van Royen and V.~F.~Weisskopf,
%``Hadron Decay Processes and the Quark Model,''
Nuovo Cim. A \textbf{50}, 617-645 (1967)
[erratum: Nuovo Cim. A \textbf{51}, 583 (1967)]
doi:10.1007/BF02823542
%724 citations counted in INSPIRE as of 24 Apr 2022

%\cite{Barbieri:1975ki}
\bibitem{Barbieri:1975ki}
R.~Barbieri, R.~Gatto, R.~Kogerler and Z.~Kunszt,
%``Meson hyperfine splittings and leptonic decays,''
Phys. Lett. B \textbf{57}, 455-459 (1975)
doi:10.1016/0370-2693(75)90267-1
%260 citations counted in INSPIRE as of 24 Apr 2022

%\cite{Celmaster:1978yz}
\bibitem{Celmaster:1978yz}
W.~Celmaster,
%``Lepton Width Suppression in Vector Mesons,''
Phys. Rev. D \textbf{19}, 1517 (1979)
doi:10.1103/PhysRevD.19.1517
%145 citations counted in INSPIRE as of 24 Apr 2022

%\cite{Bodwin:2002cfe}
\bibitem{Bodwin:2002cfe}
G.~T.~Bodwin and A.~Petrelli,
%``Order-$v^4$ corrections to $S$-wave quarkonium decay,''
Phys. Rev. D \textbf{66}, 094011 (2002)
[erratum: Phys. Rev. D \textbf{87}, no.3, 039902 (2013)]
doi:10.1103/PhysRevD.66.094011
[arXiv:hep-ph/0205210 [hep-ph]].
%157 citations counted in INSPIRE as of 24 Apr 2022

%\cite{Luke:1997ys}
\bibitem{Luke:1997ys}
M.~E.~Luke and M.~J.~Savage,
%``Power counting in dimensionally regularized NRQCD,''
Phys. Rev. D \textbf{57}, 413-423 (1998)
doi:10.1103/PhysRevD.57.413
[arXiv:hep-ph/9707313 [hep-ph]].
%93 citations counted in INSPIRE as of 24 Apr 2022

%\cite{Beneke:1997jm}
\bibitem{Beneke:1997jm}
M.~Beneke, A.~Signer and V.~A.~Smirnov,
%``Two loop correction to the leptonic decay of quarkonium,''
Phys. Rev. Lett. \textbf{80}, 2535-2538 (1998)
doi:10.1103/PhysRevLett.80.2535
[arXiv:hep-ph/9712302 [hep-ph]].
%208 citations counted in INSPIRE as of 07 Apr 2022

%\cite{Czarnecki:1997vz}
\bibitem{Czarnecki:1997vz}
A.~Czarnecki and K.~Melnikov,
%``Two loop QCD corrections to the heavy quark pair production cross-section in e+ e- annihilation near the threshold,''
Phys. Rev. Lett. \textbf{80}, 2531-2534 (1998)
doi:10.1103/PhysRevLett.80.2531
[arXiv:hep-ph/9712222 [hep-ph]].
%217 citations counted in INSPIRE as of 07 Apr 2022


%\cite{Kniehl:2006qw}
\bibitem{Kniehl:2006qw}
B.~A.~Kniehl, A.~Onishchenko, J.~H.~Piclum and M.~Steinhauser,
%``Two-loop matching coefficients for heavy quark currents,''
Phys. Lett. B \textbf{638}, 209-213 (2006)
doi:10.1016/j.physletb.2006.05.023
[arXiv:hep-ph/0604072 [hep-ph]].
%30 citations counted in INSPIRE as of 07 Apr 2022

  %\cite{Egner:2021lxd}
  \bibitem{Egner:2021lxd}
  M.~Egner, M.~Fael, J.~Piclum, K.~Schoenwald and M.~Steinhauser,
  %``Charm-quark mass effects in NRQCD matching coefficients and the leptonic decay of the \Upsilon{}(1S) meson,''
  Phys. Rev. D \textbf{104}, no.5, 054033 (2021)
  doi:10.1103/PhysRevD.104.054033
  [arXiv:2105.09332 [hep-ph]].
  %1 citations counted in INSPIRE as of 13 Apr 2022


%\cite{Marquard:2006qi}
\bibitem{Marquard:2006qi}
P.~Marquard, J.~H.~Piclum, D.~Seidel and M.~Steinhauser,
%``Fermionic corrections to the three-loop matching coefficient of the vector current,''
Nucl. Phys. B \textbf{758}, 144-160 (2006)
doi:10.1016/j.nuclphysb.2006.09.015
[arXiv:hep-ph/0607168 [hep-ph]].
%40 citations counted in INSPIRE as of 29 Apr 2022

%\cite{Marquard:2009bj}
\bibitem{Marquard:2009bj}
P.~Marquard, J.~H.~Piclum, D.~Seidel and M.~Steinhauser,
%``Completely automated computation of the heavy-fermion corrections to the three-loop matching coefficient of the vector current,''
Phys. Lett. B \textbf{678}, 269-275 (2009)
doi:10.1016/j.physletb.2009.05.070
[arXiv:0904.0920 [hep-ph]].
%25 citations counted in INSPIRE as of 29 Apr 2022

 %\cite{Marquard:2014pea}
\bibitem{Marquard:2014pea}
P.~Marquard, J.~H.~Piclum, D.~Seidel and M.~Steinhauser,
%``Three-loop matching of the vector current,''
Phys. Rev. D \textbf{89}, no.3, 034027 (2014)
doi:10.1103/PhysRevD.89.034027
[arXiv:1401.3004 [hep-ph]].
%43 citations counted in INSPIRE as of 04 Apr 2022

%\cite{Beneke:2014qea}
\bibitem{Beneke:2014qea}
M.~Beneke, Y.~Kiyo, P.~Marquard, A.~Penin, J.~Piclum, D.~Seidel and M.~Steinhauser,
%``Leptonic decay of the $\Upsilon$(1$S$) meson at third order in QCD,''
Phys. Rev. Lett. \textbf{112}, no.15, 151801 (2014)
doi:10.1103/PhysRevLett.112.151801
[arXiv:1401.3005 [hep-ph]].
%51 citations counted in INSPIRE as of 29 Apr 2022










%\cite{Liu:2017jxz}
\bibitem{Liu:2017jxz}
X.~Liu, Y.~Q.~Ma and C.~Y.~Wang,
%``A Systematic and Efficient Method to Compute Multi-loop Master Integrals,''
Phys. Lett. B \textbf{779}, 353-357 (2018)
doi:10.1016/j.physletb.2018.02.026
[arXiv:1711.09572 [hep-ph]].
%49 citations counted in INSPIRE as of 20 Feb 2022

%\cite{Liu:2020kpc}
\bibitem{Liu:2020kpc}
X.~Liu, Y.~Q.~Ma, W.~Tao and P.~Zhang,
%``Calculation of Feynman loop integration and phase-space integration via auxiliary mass flow,''
Chin. Phys. C \textbf{45}, no.1, 013115 (2021)
doi:10.1088/1674-1137/abc538
[arXiv:2009.07987 [hep-ph]].
%9 citations counted in INSPIRE as of 20 Feb 2022

%\cite{Liu:2022chg}
\bibitem{Liu:2022chg}
X.~Liu and Y.~Q.~Ma,
%``AMFlow: a Mathematica package for Feynman integrals computation via Auxiliary Mass Flow,''
[arXiv:2201.11669 [hep-ph]].
%2 citations counted in INSPIRE as of 20 Feb 2022

%\cite{Beneke:2007pj}
\bibitem{Beneke:2007pj}
M.~Beneke, Y.~Kiyo and A.~A.~Penin,
%``Ultrasoft contribution to quarkonium production and annihilation,''
Phys. Lett. B \textbf{653}, 53-59 (2007)
doi:10.1016/j.physletb.2007.06.068
[arXiv:0706.2733 [hep-ph]].
%65 citations counted in INSPIRE as of 06 Jul 2022

%\cite{Kniehl:2002yv}
\bibitem{Kniehl:2002yv}
B.~A.~Kniehl, A.~A.~Penin, M.~Steinhauser and V.~A.~Smirnov,
%``Heavy quarkonium creation and annihilation with $\mathcal{O}(\alpha_s^3  \ln(\alpha_s))$ accuracy,''
Phys. Rev. Lett. \textbf{90}, 212001 (2003)
[erratum: Phys. Rev. Lett. \textbf{91}, 139903 (2003)]
doi:10.1103/PhysRevLett.90.212001
[arXiv:hep-ph/0210161 [hep-ph]].
%62 citations counted in INSPIRE as of 06 Jul 2022

%\cite{Beneke:1997zp}
\bibitem{Beneke:1997zp}
M.~Beneke and V.~A.~Smirnov,
%``Asymptotic expansion of Feynman integrals near threshold,''
Nucl. Phys. B \textbf{522}, 321-344 (1998)
doi:10.1016/S0550-3213(98)00138-2
[arXiv:hep-ph/9711391 [hep-ph]].
%754 citations counted in INSPIRE as of 29 Apr 2022

%\cite{Kallen:1955fb}
\bibitem{Kallen:1955fb}
A.~O.~G.~Kallen and A.~Sabry,
%``Fourth order vacuum polarization,''
Kong. Dan. Vid. Sel. Mat. Fys. Med. \textbf{29}, no.17, 1-20 (1955)
doi:10.1007/978-3-319-00627-7\_93
%201 citations counted in INSPIRE as of 07 Apr 2022




%\cite{Nogueira:1991ex}
\bibitem{Nogueira:1991ex}
P.~Nogueira,
%``Automatic Feynman graph generation,''
J. Comput. Phys. \textbf{105}, 279-289 (1993)
doi:10.1006/jcph.1993.1074
%1083 citations counted in INSPIRE as of 06 Jul 2022

%\cite{Hahn:2000kx}
\bibitem{Hahn:2000kx}
T.~Hahn,
%``Generating Feynman diagrams and amplitudes with FeynArts 3,''
Comput. Phys. Commun. \textbf{140}, 418-431 (2001)
doi:10.1016/S0010-4655(01)00290-9
[arXiv:hep-ph/0012260 [hep-ph]].
%1852 citations counted in INSPIRE as of 06 Jul 2022

%\cite{Boughezal:2022nof}
\bibitem{Boughezal:2022nof}
R.~Boughezal, Y.~Huang and F.~Petriello,
%``Exploring the SMEFT at dimension-8 with Drell-Yan transverse momentum measurements,''
[arXiv:2207.01703 [hep-ph]].
%0 citations counted in INSPIRE as of 06 Jul 2022

%\cite{Mertig:1990an}
\bibitem{Mertig:1990an}
R.~Mertig, M.~Bohm and A.~Denner,
%``FEYN CALC: Computer algebraic calculation of Feynman amplitudes,''
Comput. Phys. Commun. \textbf{64}, 345-359 (1991)
doi:10.1016/0010-4655(91)90130-D
%1144 citations counted in INSPIRE as of 06 Jul 2022

%\cite{Feng:2012tk}
\bibitem{Feng:2012tk}
F.~Feng and R.~Mertig,
%``FormLink/FeynCalcFormLink : Embedding FORM in Mathematica and FeynCalc,''
[arXiv:1212.3522 [hep-ph]].
%49 citations counted in INSPIRE as of 06 Jul 2022

%\cite{Feng:2012iq}
\bibitem{Feng:2012iq}
F.~Feng,
%``$\tt{Apart}$: A Generalized Mathematica Apart Function,''
Comput. Phys. Commun. \textbf{183}, 2158-2164 (2012)
doi:10.1016/j.cpc.2012.03.025
[arXiv:1204.2314 [hep-ph]].
%78 citations counted in INSPIRE as of 06 Jul 2022

%\cite{Smirnov:2014hma}
\bibitem{Smirnov:2014hma}
A.~V.~Smirnov,
%``FIRE5: a C++ implementation of Feynman Integral REduction,''
Comput. Phys. Commun. \textbf{189}, 182-191 (2015)
doi:10.1016/j.cpc.2014.11.024
[arXiv:1408.2372 [hep-ph]].
%374 citations counted in INSPIRE as of 06 Jul 2022

%\cite{Hepp:1966eg}
\bibitem{Hepp:1966eg}
K.~Hepp,
%``Proof of the Bogolyubov-Parasiuk theorem on renormalization,''
Commun. Math. Phys. \textbf{2}, 301-326 (1966)
doi:10.1007/BF01773358
%596 citations counted in INSPIRE as of 29 Apr 2022

%\cite{Broadhurst:1991fy}
\bibitem{Broadhurst:1991fy}
D.~J.~Broadhurst, N.~Gray and K.~Schilcher,
%``Gauge invariant on-shell Z(2) in QED, QCD and the effective field theory of a static quark,''
Z. Phys. C \textbf{52}, 111-122 (1991)
doi:10.1007/BF01412333
%207 citations counted in INSPIRE as of 07 Apr 2022

%\cite{Melnikov:2000zc}
\bibitem{Melnikov:2000zc}
K.~Melnikov and T.~van Ritbergen,
%``The Three loop on-shell renormalization of QCD and QED,''
Nucl. Phys. B \textbf{591}, 515-546 (2000)
doi:10.1016/S0550-3213(00)00526-5
[arXiv:hep-ph/0005131 [hep-ph]].
%150 citations counted in INSPIRE as of 07 Apr 2022

  %\cite{Marquard:2007uj}
\bibitem{Marquard:2007uj}
P.~Marquard, L.~Mihaila, J.~H.~Piclum and M.~Steinhauser,
%``Relation between the pole and the minimally subtracted mass in dimensional regularization and dimensional reduction to three-loop order,''
Nucl. Phys. B \textbf{773}, 1-18 (2007)
doi:10.1016/j.nuclphysb.2007.03.010
[arXiv:hep-ph/0702185 [hep-ph]].
%78 citations counted in INSPIRE as of 07 Apr 2022

%\cite{Egner:2022jot}
\bibitem{Egner:2022jot}
M.~Egner, M.~Fael, F.~Lange, K.~Sch\"onwald and M.~Steinhauser,
%``Three-loop non-singlet matching coefficients for heavy quark currents,''
[arXiv:2203.11231 [hep-ph]].
%0 citations counted in INSPIRE as of 13 Apr 2022

%\cite{Choe:2003wx}
\bibitem{Choe:2003wx}
S.~Choe \textit{et al.} [QCD-TARO],
%``Quenched charmonium spectrum,''
JHEP \textbf{08}, 022 (2003)
doi:10.1088/1126-6708/2003/08/022
[arXiv:hep-lat/0307004 [hep-lat]].
%31 citations counted in INSPIRE as of 09 May 2022

%\cite{Gray:2005ur}
\bibitem{Gray:2005ur}
A.~Gray, I.~Allison, C.~T.~H.~Davies, E.~Dalgic, G.~P.~Lepage, J.~Shigemitsu and M.~Wingate,
%``The Upsilon spectrum and m(b) from full lattice QCD,''
Phys. Rev. D \textbf{72}, 094507 (2005)
doi:10.1103/PhysRevD.72.094507
[arXiv:hep-lat/0507013 [hep-lat]].
%305 citations counted in INSPIRE as of 09 May 2022

%\cite{Eichten:1995ch}
\bibitem{Eichten:1995ch}
E.~J.~Eichten and C.~Quigg,
%``Quarkonium wave functions at the origin,''
Phys. Rev. D \textbf{52}, 1726-1728 (1995)
doi:10.1103/PhysRevD.52.1726
[arXiv:hep-ph/9503356 [hep-ph]].
%391 citations counted in INSPIRE as of 01 May 2022

%\cite{Rai:2008sc}
\bibitem{Rai:2008sc}
A.~K.~Rai, B.~Patel and P.~C.~Vinodkumar,
%``Properties of $Q \bar{Q}$ mesons in non-relativistic QCD formalism,''
Phys. Rev. C \textbf{78}, 055202 (2008)
doi:10.1103/PhysRevC.78.055202
[arXiv:0810.1832 [hep-ph]].
%92 citations counted in INSPIRE as of 09 May 2022

%\cite{Chung:2020zqc}
\bibitem{Chung:2020zqc}
H.~S.~Chung,
%``$\overline {MS}$ renormalization of $S$-wave quarkonium wavefunctions at the origin,''
JHEP \textbf{12}, 065 (2020)
doi:10.1007/JHEP12(2020)065
[arXiv:2007.01737 [hep-ph]].
%8 citations counted in INSPIRE as of 09 May 2022

%\cite{Azhothkaran:2020ipl}
\bibitem{Azhothkaran:2020ipl}
B.~Azhothkaran and N.~V.~K.,
%``Decay Constants of S Wave Heavy Quarkonia,''
Int. J. Theor. Phys. \textbf{59}, no.7, 2016-2028 (2020)
doi:10.1007/s10773-020-04474-5
%1 citations counted in INSPIRE as of 06 Jul 2022

%\cite{Akbar:2015evy}
\bibitem{Akbar:2015evy}
N.~Akbar, M.~A.~Sultan, B.~Masud and F.~Akram,
%``Higher Hybrid Bottomonia in an Extended Potential Model,''
Phys. Rev. D \textbf{95}, no.7, 074018 (2017)
doi:10.1103/PhysRevD.95.074018
[arXiv:1511.03632 [hep-ph]].
%13 citations counted in INSPIRE as of 09 May 2022

%\cite{Akbar:2011jd}
\bibitem{Akbar:2011jd}
N.~Akbar, B.~Masud and S.~Noor,
%``Wave Function Based Characteristics of Hybrid Mesons,''
Eur. Phys. J. A \textbf{47}, 124 (2011)
[erratum: Eur. Phys. J. A \textbf{50}, 121 (2014)]
doi:10.1140/epja/i2011-11124-2
[arXiv:1106.3465 [hep-ph]].
%14 citations counted in INSPIRE as of 09 May 2022


%\cite{Radford:2007vd}
\bibitem{Radford:2007vd}
S.~F.~Radford and W.~W.~Repko,
%``Potential model calculations and predictions for heavy quarkonium,''
Phys. Rev. D \textbf{75}, 074031 (2007)
doi:10.1103/PhysRevD.75.074031
[arXiv:hep-ph/0701117 [hep-ph]].
%116 citations counted in INSPIRE as of 06 Jul 2022

%\cite{Bodwin:2007fz}
\bibitem{Bodwin:2007fz}
G.~T.~Bodwin, H.~S.~Chung, D.~Kang, J.~Lee and C.~Yu,
%``Improved determination of color-singlet nonrelativistic QCD matrix elements for S-wave charmonium,''
Phys. Rev. D \textbf{77}, 094017 (2008)
doi:10.1103/PhysRevD.77.094017
[arXiv:0710.0994 [hep-ph]].
%121 citations counted in INSPIRE as of 28 Jul 2022

%\cite{Jegerlehner:2011mw}
\bibitem{Jegerlehner:2011mw}
F.~Jegerlehner,
%``Electroweak effective couplings for future precision experiments,''
Nuovo Cim. C \textbf{034S1}, 31-40 (2011)
doi:10.1393/ncc/i2011-11011-0
[arXiv:1107.4683 [hep-ph]].
%90 citations counted in INSPIRE as of 23 Jun 2022

%\cite{Herren:2017osy}
\bibitem{Herren:2017osy}
F.~Herren and M.~Steinhauser,
%``Version 3 of RunDec and CRunDec,''
Comput. Phys. Commun. \textbf{224}, 333-345 (2018)
doi:10.1016/j.cpc.2017.11.014
[arXiv:1703.03751 [hep-ph]].
%122 citations counted in INSPIRE as of 01 May 2022









\end{thebibliography}
\end{document}